\documentclass[american,raggedbottom]{revtex4-2}
\usepackage[T1]{fontenc}
\usepackage[utf8]{inputenc}
\usepackage{xcolor}
\usepackage{colortbl}
\usepackage{babel}
\usepackage{array}
\usepackage{mathtools}
\usepackage{enumitem}
\usepackage{multirow}
\usepackage{dsfont}
\usepackage{amsmath}
\usepackage{graphicx}
\usepackage[pdfusetitle,
 bookmarks=true,bookmarksnumbered=false,bookmarksopen=false,
 breaklinks=false,pdfborder={0 0 0},pdfborderstyle={},backref=false,colorlinks=true]
 {hyperref}
\hypersetup{
 urlcolor=blue, citecolor=blue, linkcolor=blue}

\makeatletter

\providecommand{\tabularnewline}{\\}

\usepackage[justification=raggedright,singlelinecheck=false]{caption}

\makeatother

\begin{document}
\title{The role of different nonlinearities and potential vorticity conservation
in two-dimensional fluid ITG models}
\author{G. Paramasivam}
\thanks{Corresponding author}
\email{giridharan.paramasivam@lpp.polytechnique.fr}

\author{Ö. D. Gürcan}
\affiliation{Laboratoire de Physique des Plasmas, CNRS, Ecole Polytechnique, Sorbonne
Université, Université Paris-Saclay, Observatoire de Paris, F-91120
Palaiseau, France}
\begin{abstract}
The nature of turbulent energy cascade of simple two-dimensional fluid
models of ion temperature gradient driven turbulence is studied in
detail. Notably, it is observed that a minimal two-field model of
toroidal ITG behaves qualitatively differently with or without the
diamagnetic nonlinearity. In its absence, the zonal flows always dominate
and the system never reaches a high-transport state. In contrast,
when this term is included, zonal flows dominate only near marginality,
while away from it, an inverse cascade with high levels of transport
is observed, requiring large-scale dissipation (hypoviscosity) to
saturate and hyperviscosity to regularize small scale instability
associated with this nonlinear term. However introducing such a term,
together with the existence of the curvature term, breaks potential
vorticity conservation, which is one of the key symmetries of drift-wave
turbulence. This can be remedied by considering a more complete model
that retains higher-order terms in the pressure equation. This form
of the model, with four nonlinearities, conserves potential vorticity
and behaves similarly to the original model, requiring hyperviscosity
to saturate since the added nonlinearity generates small scale instability
also for the pressure equation. To characterize the roles of potential
vorticity conservation and higher-order terms in the pressure equation,
the behavior of both the standard ITG system, and the potential vorticity
conserving system, is studied by analyzing their spectra, turbulent
cascades, and sensitivity to viscosity. Finally, the direction of
turbulent cascade due to the different nonlinearities is investigated
by examining their contributions to the spectral energy transfer and
by considering the triadic instability assumption for each of these
nonlinearities separately.
\end{abstract}
\maketitle

\section{Introduction\label{sec:Introduction}}

Turbulent transport is important in magnetized fusion plasmas, since
neoclassical theory fails to explain the observed transport levels
at low collisionality~\citep{Groebner_1986,Haskey_2022}. The dominant
instability responsible for ion heat transport in low-$\beta$ plasma
cores is considered to be the ion temperature gradient driven (ITG)
mode, usually coexisting with the trapped electron mode (TEM)~\citep{Garbet_2004}.
The ITG instability~\citep{Rudakov_1961,Coppi_1967,Guzdar_1983}
provides the free energy source for turbulent fluctuations at the
scale of the ion gyroradius, $\rho_{i}$, and it is these micro-scale
fluctuations that generate turbulent transport in fusion plasmas~\citep{Horton_1981,Horton_1999}.

The toroidal ITG instability considered here is somewhat analogous
to Rayleigh-Bénard instability, with magnetic curvature playing the
role of gravity, while the temperature gradient provides the free
energy for the turbulent motions of the fluid. It is a reactive instability
that grows even in the collisionless limit, for instance with adiabatic
electrons, in contrast to dissipative instabilities such as the standard
drift-wave instability. One might, a priori, expect a linear analysis
to accurately predict the onset of turbulent transport in such a system,
once the temperature gradient exceeds a critical value. However, such
an approach neglects the effect of zonal flows, which can suppress
transport resulting in an intermediate state dominated by zonal flows~\citep{Dimits_2000},
somewhat analogous to the Bussé balloon in Rayleigh-Bénard convection~\citep{Busse_1978,St-Onge_Krommes_2017}.
Significant turbulence and high transport emerge only after crossing
a second, higher nonlinear threshold ($\kappa_{T,\mathrm{nl}}$).
This nonlinear shift in the critical gradient is called the `Dimits
Shift’ and is essential for describing the suppression and onset of
transport. Therefore, even an absolutely minimal model of ITG must
be capable of capturing this behavior, if it is to be useful for studying
the self-organization of plasma turbulence.

Bifurcations of this kind, from a zonal-flow-dominated state to a
high-transport state, have also been observed in gradient-driven simulations
of resistive drift-wave turbulence~\citep{Numata_2007} along with
a hysteresis loop for the classical~\citep{Guillon_2025} and gyrofluid~\citep{Grander_2024}
Hasegawa-Wakatani systems. Similarly, hysteresis about the nonlinear
threshold for ITG turbulence using gradient driven gyrokinetic simulations
has also been recently reported~\citep{Marquant_2026}. Ideally,
one would like to recover the hysteresis associated with the transition
using the simple fluid models with high resolution simulations. However,
the additional flexibility provided by using fluid models makes the
choice of model an essential issue. Therefore, here we focus on the
physics of the system when it is in either of the states that it presents
and try to characterize the effects of different nonlinearities, leaving
the investigation of possible hysteresis behavior to a future study.

The state of the art in the study of plasma turbulence is the use
of global, electromagnetic, gyrokinetic simulations. However, the
additional cost to run and analyze these models does not justify the
gain in understanding, especially if one is interested in the nonlinear
interplay and feedback loops that such a system provides. Therefore,
to study the nonlinear behavior of the ITG system in a light and efficient
manner, one may opt for fluid and gyrofluid models~\citep{Brizard_1992,Dorland_1993,Beer_1996_gyrofluid},
which allow for cheaper 2D simulations with much higher spatial resolution
instead of costly and usually lower-resolution simulations in 5D.
In doing so, one loses various linear and some nonlinear mechanisms
such as wave-particle resonances, Landau damping, trapped or energetic
particles etc., unless they are modeled in some way using approximate
closures~\citep{Hammett_1990,Waltz_1992,Beer_1996_trapped,Staebler_2005}
or by considering different classes of particles as separate species~\citep{Weiland_1989,Nordman_1990}.
As long as these features are not essential for the mechanism that
one wants to study, using minimal 2D fluid models, as we discuss in
detail in this paper, can be justified.

The study of ITG turbulence using 2D fluid models is nothing new~\citep{Lee_1986,Nordman_1989,Hamaguchi_1990}.
Previous studies have investigated the effects of finite-Larmor-radius
(FLR) corrections~\citep{Kim_1993} and the generation of zonal flows~\citep{Anderson_2002}.
A Hamiltonian formulation of an FLR fluid ITG model has also been
used to study conservation laws and coherent structures~\citep{Waelbroeck_2004}.
The roles of the polarization-drift and convective ion-heat nonlinearities
in the evolution and saturation of turbulence have also been investigated~\citep{Sandberg_2006}.

However, accurately capturing the physics of the nonlinear threshold
for the onset of turbulent transport remains a challenge for reduced
fluid models, although some fluid models have reproduced the nonlinear
upshift of the critical temperature gradient~\citep{Weiland_2004}.
It has recently been suggested that the diamagnetic nonlinearity~\citep{Horton_1980,Brizard_1992,Horton_1992,Smolyakov_1998},
and the associated diamagnetic stress, may be the key for the destabilization
of the zonal flows at a critical nonlinear threshold~\citep{Ivanov_2020}.
It has also been noted that the stress associated with the diamagnetic
nonlinearity might be comparable to or larger than the Reynolds stress~\citep{Smolyakov_2000,Madsen_2017,Sarazin_2021,Dif-Pradalier_2022}.
In particular, it is suggested in reference~\citealp{Sarazin_2021}
that, in the core, the diamagnetic stress can exceed the Reynolds
stress by a factor of two, contradicting earlier reports~\citep{Dimits_2007}
that may not have allowed simulations to reach full saturation. Therefore,
to be sure that the nonlinear results obtained are reliable, we run
the simulations for a sufficiently long duration: $T=1000\gamma^{-1}_{\mathrm{max}}$,
where $\gamma_{\mathrm{max}}$ is the maximum linear growth rate for
a given set of parameters for the system.

However, for example, the model proposed in reference~\citealp{Ivanov_2020}
leads to blow up beyond the zonal-flow-dominated regime and the followup~\citep{Ivanov_2022}
resolves the issue by performing 3D simulations. However, as we discuss
in this paper, one can stick to a 2D model and use hypoviscosity to
achieve a saturated turbulent state as is conventionally done in simulations
of two-dimensional homogeneous-isotropic turbulence to avoid condensate
formation due to inverse cascade of energy~\citep{Kraichnan_1967,Alexakis_2018}.
However, in our case, even with hypoviscosity, the evolution of the
system is found to be bursty with the formation of streamer dominated
states and small-scale numerical instabilities (section~\ref{sec:Nonlinear-simulations}),
which suggests that the use of hyperviscosity may also be necessary
in place of Laplacian viscosity especially since we lack parallel
damping phenomena such as Landau damping.

In this paper, to distinguish the two main models that we consider,
we refer to the equations analogous to reference~\citealp{Ivanov_2020}
as the standard ITG (ST-ITG) models. These models have FLR terms only
in the potential equation but not the pressure equation. This is good
enough to obtain the physics of zonal flow destabilization at a critical
temperature gradient. However, these models do not conserve potential
vorticity~\citep{Gurcan_2015} due to the omission of FLR terms in
the pressure equation. Since the radial flux of potential vorticity
is linked to zonal flow formation in Charney-Hasegawa-Mima (CHM) and
equivalent systems, we would like to construct an ITG model that conserves
potential vorticity as well. It has been noted that for these systems,
the potential vorticity is simply the gyrocenter density~\citep{Hahm_2024}.
However, the introduction of pressure modifies this simple picture.
So, we propose a potential vorticity conserving fluid model for ITG,
based on the potential vorticity given in reference~\citealp{Gurcan_2015}
and making sure we conserve it by retaining some higher-order terms
in the pressure equation, which are normally dropped. These terms
facilitate the saturation of the heat flux while preserving the zonal-flow-dominated
state to high-transport state transition.

In this work, we study two fluid ITG models that exhibit the Dimits
shift physics as described in section~\ref{sec:2D-Fluid-ITG}. We
first present the standard ITG model (ST-ITG) analogous to the model
in reference~\citealp{Ivanov_2020} and then the potential vorticity
conserving fluid ITG model (PV-ITG) that we propose. We study the
effects of the FLR term on the growth rate and the dependence of turbulent
diffusivity estimates on the hypoviscosity in section~\ref{sec:Linear-Analysis}.
Furthermore, we explain the necessity of using both hypoviscosity
and hyperviscosity in simulations of such models along with a comparison
in the evolution of the zonal kinetic energy fraction, box-averaged
heat flux and zonal velocity profile of the two models in section~\ref{sec:Nonlinear-simulations}.
In section~\ref{sec:Role-of-flr-nl}, we study the spectra, spectral
fluxes and the cascade direction via Waleffe's ``Instability assumption''
approach~\citep{Waleffe_1992,Waleffe_1993}. We explore the energy
cascading behaviors of the two models and the contribution of the
FLR nonlinearities. In section~\ref{sec:Viscous-dissipation}, we
note the role of the hyperviscosity coefficient on the zonal flow
levels of the ST-ITG model, and show that the PV-ITG model is more
resilient in its zonal flow levels when the hyperviscosity is lowered
compared to the ST-ITG model. Finally we conclude in section~\ref{sec:Conclusion}
with key takeaways of this study together with a possible 3D potential
vorticity conserving fluid ITG model.

\section{2D Fluid ITG Models\label{sec:2D-Fluid-ITG}}

In order to construct a minimal model of ITG turbulence we consider
two fields: the pressure and the electric potential. We consider the
low-$\beta$ limit ($\beta\sim\epsilon^{2}$ where $\epsilon\coloneqq\rho_{i0}/L_{n}$),
which also justifies the omission of the parallel vector potential
$A_{\parallel}$, thereby restricting the model to an electrostatic
formulation. The fluctuations and independent variables are written
in non-dimensionalized form as
\begin{equation}
\begin{aligned}n' & =\frac{n_{1}}{n_{i0}}\,\text{,} & P' & =\frac{P_{i1}}{n_{i0}T_{i0}}\,\text{,} & \phi' & =\frac{e\phi_{1}}{T_{i0}}\,\text{,}\\
t' & =\Omega_{i0}t\,\text{,} & l' & =\frac{l}{\rho_{i0}}\,\text{,} & \boldsymbol{k}' & =\boldsymbol{k}\rho_{i0}\,\text{,}\\
\kappa_{n} & =\frac{\rho_{i0}}{L_{n}}\,\text{,} & \kappa_{T} & =\frac{\rho_{i0}}{L_{T}}\,\text{,} & \kappa_{B} & =\frac{2\rho_{i0}}{R}\,\text{,}
\end{aligned}
\end{equation}

\noindent where $n_{i0}$ and $T_{i0}$ are the equilibrium density
and ion temperature, respectively, $\Omega_{i0}=eB_{0}/m_{i}$ is
the ion gyrofrequency and $\rho_{i0}=\sqrt{T_{i0}/m_{i}}/\Omega_{i0}$
is the ion gyroradius. Here, $n'$ is the normalized density fluctuation,
$P'$ is the normalized pressure fluctuation, and $\phi'$ is the
normalized electric potential.

The background profiles are characterized by their characteristic
inverse gradient length scales. The inverse density and ion temperature
gradient length scales are defined as $L^{-1}_{n}=-\partial_{x}n_{i0}/n_{i0}$
and $L^{-1}_{T}=-\partial_{x}T_{i0}/T_{i0}$, respectively. The term
$2R^{-1}$ is the sum of the magnetic curvature ($R^{-1}$) and the
inverse magnetic field gradient length scale ($L^{-1}_{B}$) which
are approximately equal in the low-$\beta$ limit. Multiplying these
inverse gradient length scales by the ion Larmor radius, $\rho_{i0}$,
yields the dimensionless parameters $\kappa_{n}$, $\kappa_{T}$,
and $\kappa_{B}$. We use the standard gyrokinetic ordering~\citep{Frieman_1982}
and a large-aspect-ratio ordering $\kappa_{B}/\kappa_{n,T}\sim\epsilon$.
Furthermore, in this paper, we omit the prime of the fluctuation variables
for simplicity, and drop the `$\perp$' from $\boldsymbol{k}_{\perp}$
and simply use $\boldsymbol{k}$. However the individual components
still come with subscripts: $k_{x}$ and $k_{y}$.

\subsection{Standard ITG model (ST-ITG)\label{subsec:ST-ITG-model}}

Following reference~\citealp{Brizard_1992}, and simplifying the
resulting system further by considering the electrostatic limit, adiabatic
electrons, large aspect ratio, and small finite Larmor radius (FLR),
we obtain the 2D toroidal ($k_{\parallel}=0)$ ITG system analogous
to that in reference~\citealp{Ivanov_2020},
\begin{align}
\partial_{t}P+\{\phi,P\}+(\kappa_{n}+\kappa_{T})\partial_{y}\phi & =D_{P}\,\text{,}\label{eq:itg2d-1}\\
\partial_{t}(\tau\widetilde{\phi}-\nabla^{2}\phi)+\{\phi,(\tau\widetilde{\phi}-\nabla^{2}\phi)\}+\boldsymbol{\nabla}\cdot\{\boldsymbol{\nabla}\phi,P\}+\kappa_{n}\partial_{y}\phi+(\kappa_{n}+\kappa_{T})\partial_{y}\nabla^{2}\phi-\kappa_{B}\partial_{y}P & =D_{\phi}\,\text{,}\label{eq:itg2d-2}
\end{align}

\noindent where the Poisson bracket $\{\phi,P\}$ denotes advection
of $P$ by the $\mathrm{E}\times\mathrm{B}$ velocity, $\boldsymbol{v}_{E}=\boldsymbol{z}\times\boldsymbol{\nabla}\phi$,
so that the term $(\kappa_{n}+\kappa_{T})\partial_{y}\phi$ represents
the advection of the background pressure gradient that provides free
energy to turbulence.

We assume a quasineutral plasma with adiabatic electrons (i.e. no
particle transport), $n_{i}=n_{e}=\tau\widetilde{\phi}$ where $\tau=T_{i0}/T_{e0}$
is the ion-to-electron temperature ratio. Here, $\widetilde{\phi}\coloneqq\phi-\overline{\phi}$
is the non-zonal component of $\phi$ and $\overline{\phi}=\int\mathrm{d}y~\phi/\int\mathrm{d}y$
is the zonal average of $\phi$ . The $\mathrm{E}\times\mathrm{B}$
vorticity {[}i.e. $(\boldsymbol{\nabla}\times\boldsymbol{v}_{E})_{z}=\nabla^{2}\phi${]}
equation is obtained from the quasineutrality condition, and subtracting
it from the density equation gives the equation for $\tau\widetilde{\phi}-\nabla^{2}\phi$
{[}equation~(\ref{eq:itg2d-2}){]}.

$\boldsymbol{\nabla}\cdot\{\boldsymbol{\nabla}\phi,P\}$ is an additional
FLR nonlinearity originating from the FLR contribution to the divergence
of the ion polarization current~\citep{Brizard_1992}. It represents
the diamagnetic ``advection'' of vorticity and is important for
destabilizing the zonal flows in the zonal-flow-dominated regime.
The Boussinesq approximation is implicit in the polarization current
contribution, since, under the standard small-amplitude gyrokinetic
ordering, the neglected density-dependent correction is of order $\mathcal{O}(\epsilon^{3})$,
whereas the retained diamagnetic nonlinearity is of order $\mathcal{O}(\epsilon^{2})$.
In contrast, the density factor in the diamagnetic-advection contribution
cancels the inverse-density dependence of the diamagnetic drift. The
term $\kappa_{n}\partial_{y}\phi$ is the $\mathrm{E}\times\mathrm{B}$
advection of the background density gradient and $(\kappa_{n}+\kappa_{T})\partial_{y}\nabla^{2}\phi$
is the FLR term associated with the diamagnetic drift due to the background
pressure gradient. Note that when a constant gradient background pressure
term is used in the diamagnetic nonlinearity, as in $\boldsymbol{\nabla}\cdot\{\boldsymbol{\nabla}\phi,P_{0}(x)\}$,
it would give this term. Finally, the cross term $\kappa_{B}\partial_{y}P$
represents the magnetic drift of $P$, which includes both the curvature
and the $\nabla B$ drifts. This term is important because it couples
the pressure equation to the potential equation and, above a certain
$\kappa_{T}$, is responsible for reinforcing the potential fluctuation
leading to the toroidal ITG instability. The curvature drift facilitates
the pressure gradient to perform `work' on the plasma.

The quantities that are conserved in the absence of background gradients
and dissipation are $\langle P^{2}\rangle/2$, $\langle\tau\widetilde{\phi}^{2}+(\boldsymbol{\nabla}\phi)^{2}\rangle/2$,
and $\langle\tau(\widetilde{\phi}+P)^{2}+(\boldsymbol{\nabla}\phi+\boldsymbol{\nabla}P)^{2}\rangle/2$,
representing the internal energy from pressure ($E_{P}$), the total
energy ($E_{\mathrm{tot}}$), and what we refer to as the generalized
total energy ($G_{\mathrm{tot}}$) respectively. Here, the angle brackets
$\langle\cdots\rangle$ denote the normalized two-dimensional spatial
average (box average). Furthermore, the potential and the kinetic
energies can be defined as $\langle\tau\widetilde{\phi}^{2}\rangle/2$
and $\langle(\boldsymbol{\nabla}\phi)^{2}\rangle/2$ respectively.
The third invariant, $G_{\mathrm{tot}}$, is obtained by computing
the cross terms of the fields (derivation shown in appendix~\ref{appsec:3rd-Conservation-law}).
The square of sums in $G_{\mathrm{tot}}$ results in the phase between
the potential and the pressure fluctuations as an important quantity
in the system's evolution and cascading in addition to their amplitudes.
The first two conserved quantities can be combined to produce $\langle\kappa_{B}P^{2}-(\kappa_{n}+\kappa_{T})[\tau\widetilde{\phi}^{2}+(\boldsymbol{\nabla}\phi)^{2}]\rangle$,
which is conserved in the dissipationless limit without requiring
the limit of vanishing gradients.

The diamagnetic nonlinearity results in the non-conservation of both
what we call the drift-wave potential vorticity, $\nabla^{2}\phi-\tau\widetilde{\phi}$,
and the drift-wave potential enstrophy, $\langle(\tau\widetilde{\phi}-\nabla^{2}\phi)^{2}\rangle/2$,
in the limit of vanishing gradients and dissipation. This makes the
conserved quantities and the cascading behavior of the ST-ITG system
different in comparison to other simpler fluid ITG systems that don't
have the diamagnetic nonlinearity~\citep{Horton_1981,Hu_1997}. Hence,
the Fjortoft argument that the conservation of the higher-order drift-wave
potential enstrophy leads to an inverse cascade of energy does not
hold for this system (also see \ref{subsec:Instability-assumption}).

We use hyperviscosities, $D\nabla^{6}P$ and $\nu\nabla^{6}(\tau\widetilde{\phi}-\nabla^{2}\phi)$,
to damp small scales to regularize the numerical system. Since for
large temperature gradients beyond the zonal-flow-dominated regime,
the system displays a tendency towards inverse cascade, which results
in the accumulation of energy in large-scale condensates (in the form
of box sized streamers), we also impose a large scale damping by introducing
hypoviscous dissipation, $H\nabla^{-4}\widetilde{P}$ and $H\nabla^{-4}(\tau\widetilde{\phi}-\nabla^{2}\widetilde{\phi})$,
which is somewhat common in 2D simulations~\citep{Xiao_2009,Bracco_2010,Jha_2025}
to allow for steady state saturation of turbulence. This large-scale
damping is phenomenological and may represent the net effect of physical
processes absent from our two-dimensional fluid model, although the
operator is not derived from any specific physical mechanism. However,
we choose to exclude zonal flows from this large-scale damping, as
they do not generate any transport (in any case, the hypoviscosity
is only relevant when the zonal flows are not dominant). In principle,
one could use different hypoviscosity coefficients; however, since
we choose them to be as small as numerically feasible, we fix them
to be the same. This explicitly gives $D_{P}=D\nabla^{6}P-H\nabla^{-4}\widetilde{P}$
and $D_{\phi}=\nu\nabla^{6}(\tau\widetilde{\phi}-\nabla^{2}\phi)-H\nabla^{-4}(\tau\widetilde{\phi}-\nabla^{2}\widetilde{\phi})$.

\subsection{ITG model conserving potential vorticity (PV-ITG)\label{subsec:PV-ITG-model}}

In drift-wave turbulence, the inhomogeneous mixing of the potential
vorticity is linked to zonal flow formation. This is due to what is
called Taylor's identity, which links the radial transport of potential
vorticity to the Reynolds stress, whose radial derivative drives the
zonal flows. The basic picture holds even in a gyrokinetic formulation
of drift-wave turbulence equivalent to the Charney-Hasegawa-Mima (CHM)
system~\citep{McDevitt_2010,Hahm_2024}.

However for ITG, unlike simple drift waves, the gyrocenter pressure
enters the system of equations alongside the gyrocenter density, and
therefore, the potential vorticity should be modified to also include
the gyrocenter pressure. It is thus worth considering how an ITG system
that conserves potential vorticity can be constructed. To do so, we
consider modifications on the pressure equation, that would somehow
restore the conservation of a potential vorticity that is defined
as $q=\nabla^{2}\phi-n+P/\Gamma$ as suggested in reference~\citealp{Gurcan_2015}.
This gives
\begin{align}
\begin{aligned}\partial_{t}P+\{\phi,P\}+\Gamma\boldsymbol{\nabla}\cdot\{\boldsymbol{\nabla}\phi,P\} & +(\kappa_{n}+\kappa_{T})\partial_{y}\phi+\Gamma(\kappa_{n}+\kappa_{T})\partial_{y}\nabla^{2}\phi-\Gamma\kappa_{B}\partial_{y}P\end{aligned}
 & =D_{P}\,\text{,}\label{eq:itg2d_pv-1}\\
\partial_{t}(\tau\widetilde{\phi}-\nabla^{2}\phi)+\{\phi,(\tau\widetilde{\phi}-\nabla^{2}\phi)\}+\boldsymbol{\nabla}\cdot\{\boldsymbol{\nabla}\phi,P\}+\kappa_{n}\partial_{y}\phi+(\kappa_{n}+\kappa_{T})\partial_{y}\nabla^{2}\phi-\kappa_{B}\partial_{y}P & =D_{\phi}\,\text{,}\label{eq:itg2d_pv-2}
\end{align}
which conserves the generalized potential vorticity defined as $\text{PV}\coloneqq\nabla^{2}\phi-\tau\widetilde{\phi}+P/\Gamma+(\kappa_{n}-[\kappa_{n}+\kappa_{T}]/\Gamma)x$
in the dissipationless limit where $q=\nabla^{2}\phi-\tau\widetilde{\phi}+P/\Gamma$
is the perturbed part. Note that $\tau\widetilde{\phi}$ would simply
be $\widetilde{n}_{e}$ if the electrons were not assumed to be adiabatic.
Moreover, $\kappa_{n}-[\kappa_{n}+\kappa_{T}]/\Gamma$ is the term
in the expression of $\text{PV}$ corresponding to the background
profile of ion density and ion pressure. This introduces three higher-order
terms in the pressure equation, all of which are multiplied by the
adiabatic index $\Gamma$: the FLR nonlinearity, $\Gamma\boldsymbol{\nabla}\cdot\{\boldsymbol{\nabla}\phi,P\}$;
a linear FLR term, $\Gamma(\kappa_{n}+\kappa_{T})\partial_{y}\nabla^{2}\phi$;
and a curvature term, $-\Gamma\kappa_{B}\partial_{y}P$, which is
higher order due to the large aspect ratio ordering, $\kappa_{B}/\kappa_{n,T}\sim\epsilon\ll1$.
Note that all of these terms are present in the general form of Brizard's
gyrofluid equations~\citep{Brizard_1992}, but are dropped in the
usual derivation. Keeping them allows us to conserve potential vorticity,
which the original gyrokinetic system somehow preserved at least in
the drift-wave limit. However, not all $\mathcal{\ensuremath{O}}(\kappa_{B}/\kappa_{n,T})$
terms need to be retained; one could still neglect $\kappa_{B}$ with
respect to $\kappa_{n,T}$ when they appear in additive form in the
coefficient of a linear term. For example $(\kappa_{n}-\kappa_{B})\partial_{y}\phi\approx\kappa_{n}\partial_{y}\phi,$
as we have done to arrive at the PV-ITG system.

In contrast to the generalized PV considered here, the long-wavelength
gyrofluid PV derived in reference~\citealp{Madsen_2015} from a full-$F$
model contains the ion pressure through diamagnetic polarization corrections.
Our generalized PV also differs from the magnetic potential vorticity
derived in the cold-ion limit in reference~\citealp{Hahm_2024},
which explicitly accounts for spatial variations in the magnetic field.
In our local formulation, the terms proportional to $\kappa_{B}\partial_{y}P$
cancel in the PV evolution equation, while the contribution $-\kappa_{B}\partial_{y}\phi$
in the potential equation is neglected under the assumed large-aspect-ratio
ordering. Similarly, when using the PV expression of reference~\citealp{Gurcan_2015}
to motivate the form of our generalized PV, we retain the equilibrium
density and pressure gradients while neglecting contributions arising
from the spatial dependence of the magnetic field. Consequently, the
present model does not reproduce the full magnetic-PV structure associated
with a spatially varying magnetic field.

We use exactly the same dissipation operators as in the ST-ITG system,
$D_{P}=D\nabla^{6}P-H\nabla^{-4}\widetilde{P}$ and $D_{\phi}=\nu\nabla^{6}(\tau\widetilde{\phi}-\nabla^{2}\phi)-H\nabla^{-4}(\tau\widetilde{\phi}-\nabla^{2}\widetilde{\phi})$.
The hyperviscous terms damp small-scale modes, while the hypoviscous
terms suppress the accumulation of energy at large scales due to the
inverse cascade in the high-transport regime. As in the ST-ITG system,
the hypoviscosity acts exclusively on nonzonal modes, with the same
coefficient $H$ used in both equations. Although the dissipation
coefficients can be chosen independently for the two models, we use
identical coefficients in both models at a given value of $\kappa_{T}$.
Moreover, we also set $D=\nu$.

Note that the system would also conserve potential vorticity if one
drops all the FLR terms, but that would result in the loss of zonal
flow destabilization physics that we would like to retain. The message
is that if we keep some FLR terms to retain some interesting physics,
we may end up breaking a mixed conservation law, and the way to restore
it may be either to drop those terms and the physics with them, or
to keep the terms in a consistent way in both equations, so that the
conservation law can be recovered.

The PV-ITG system {[}equations~(\ref{eq:itg2d_pv-1}) and~(\ref{eq:itg2d_pv-2}){]}
also conserves total energy, $E_{\mathrm{tot}}=\langle\tau\widetilde{\phi}^{2}+(\boldsymbol{\nabla}\phi)^{2}\rangle/2$.
However, it no longer conserves $E_{P}=\langle P^{2}\rangle/2$, due
to the FLR nonlinearity in the pressure equation. Instead, it conserves
potential enstrophy, $W=\langle(\nabla^{2}\phi-\tau\widetilde{\phi}+P/\Gamma)^{2}\rangle/2$,
\begin{equation}
\partial_{t}\langle(\nabla^{2}\phi-\tau\widetilde{\phi}+P/\Gamma)^{2}\rangle/2=[\kappa_{n}-(\kappa_{n}+\kappa_{T})/\Gamma]\langle(\nabla^{2}\phi-\tau\widetilde{\phi}+P/\Gamma)\partial_{y}\phi\rangle-\nu\langle(\boldsymbol{\nabla}^{3}q)^{2}\rangle-H\langle(\nabla^{-2}\widetilde{q})^{2}\rangle,\label{eq:potential_enstrophy_conservation}
\end{equation}
where we have assumed $D=\nu$. Note that the PV-ITG system does not
conserve the drift-wave potential enstrophy, $\langle(\tau\widetilde{\phi}-\nabla^{2}\phi)^{2}\rangle/2$,
separately. Although the potential enstrophy is a quadratic invariant,
it is not purely a function of the potential like the total energy.
Instead, it contains, in addition to the drift-wave potential enstrophy,
a cross-term between $\nabla^{2}\phi-\tau\widetilde{\phi}$ and $P/\Gamma$
as well as $E_{P}$. Therefore, energy can be transferred to small-scale
modes without potential enstrophy conservation requiring a simultaneous
transfer to low-wavenumber modes as in 2D Navier-Stokes turbulence~\citep{Fjortoft_1953}.
See section~\ref{subsec:Instability-assumption} for further discussion.

Note that while our equations have a very similar structure, the pressure
equation~(72) of reference~\citealp{Brizard_1992}, after substituting
the vorticity equation~(63), has an extra $-\Gamma\boldsymbol{\nabla}\cdot\{\boldsymbol{\nabla}\phi,\tau\widetilde{\phi}\}$
and $-\Gamma\kappa_{n}\partial_{y}\nabla^{2}\phi$ in the LHS of the
pressure equation~(\ref{eq:itg2d_pv-1}) and this would not conserve
$\nabla^{2}\phi-\tau\widetilde{\phi}+P/\Gamma+(\kappa_{n}-[\kappa_{n}+\kappa_{T}]/\Gamma)x$
in the dissipationless limit. This discrepancy might be due to the
use of pressure in the FLR terms where in fact temperature should
have been used~\citep{Snyder_1999}. Nevertheless, if the overall
motivation is to conserve potential vorticity, then the above system
is the obvious solution.

\section{Linear Analysis\label{sec:Linear-Analysis}}

The dispersion relation of the ST-ITG system {[}equations~(\ref{eq:itg2d-1})
and~(\ref{eq:itg2d-2}){]} can be written as
\begin{equation}
\begin{aligned}\omega^{2}_{\boldsymbol{k}}(\tau+k^{2}) & -\omega_{\boldsymbol{k}}[-2iC_{k}k^{2}(\tau+k^{2})+\kappa_{n}k_{y}-(\kappa_{n}+\kappa_{T})k_{y}k^{2}]\\
 & -iC_{k}k^{2}[\kappa_{n}k_{y}-(\kappa_{n}+\kappa_{T})k_{y}k^{2}-iC_{k}k^{2}(\tau+k^{2})]+\kappa_{B}(\kappa_{n}+\kappa_{T})k^{2}_{y}=0\,\text{,}
\end{aligned}
\label{eq:dispersion_relation}
\end{equation}

\noindent while that of the PV-ITG system {[}equations~(\ref{eq:itg2d_pv-1})
and~(\ref{eq:itg2d_pv-2}){]} is slightly different, taking the form
\begin{equation}
\begin{aligned}\omega^{2}_{\boldsymbol{k}}(\tau+k^{2}) & -\omega_{\boldsymbol{k}}[-\Gamma\kappa_{B}k_{y}-2iC_{k}k^{2}(\tau+k^{2})+\kappa_{n}k_{y}-(\kappa_{n}+\kappa_{T})k_{y}k^{2}]\\
 & -[\Gamma\kappa_{B}k_{y}+iC_{k}k^{2}][\kappa_{n}k_{y}-(\kappa_{n}+\kappa_{T})k_{y}k^{2}-iC_{k}k^{2}(\tau+k^{2})]+\kappa_{B}[(\kappa_{n}+\kappa_{T})-\Gamma(\kappa_{n}+\kappa_{T})k^{2}]k^{2}_{y}=0\,\text{,}
\end{aligned}
\label{eq:dipersion_relation_PV}
\end{equation}

\noindent where $C_{k}=\nu k^{4}+\mathds{1}_{k_{y}\neq0}Hk^{-6}$
and $\omega_{\boldsymbol{k}}=\omega_{\boldsymbol{k}r}+i\gamma_{\boldsymbol{k}}$
. Here $\mathds{1}_{k_{y}\neq0}$ is the indicator function that maps
$k_{y}\neq0$ to $1$ and $k_{y}=0$ to $0$. Hence the Fourier transform
of $\widetilde{\phi}$ is $\mathcal{F}_{\boldsymbol{k}}(\widetilde{\phi})=\widetilde{\phi}_{\boldsymbol{k}}=\mathds{1}_{k_{y}\neq0}\phi_{\boldsymbol{k}}$.
The dispersion relation is quadratic in $\omega_{\boldsymbol{k}}$
and has two corresponding solutions for $\gamma_{\boldsymbol{k}}$,
from which we pick the unstable branch. The FLR terms in the dispersion
relation ($\propto[\kappa_{n}+\kappa_{T}]k^{2}k_{y}$) play an important
role in determining the functional form of the growth rate at large
wavenumbers. Note also that in the dissipationless limit, setting
$\kappa_{B}=0$ yields a stable ion drift wave ($\gamma_{\boldsymbol{k}}=0$)
with real frequency $\omega_{\boldsymbol{k}r}=(\kappa_{n}k_{y}-[\kappa_{n}+\kappa_{T}]k_{y}k^{2})/(\tau+k^{2})$.

\begin{figure*}
\begin{centering}
\includegraphics{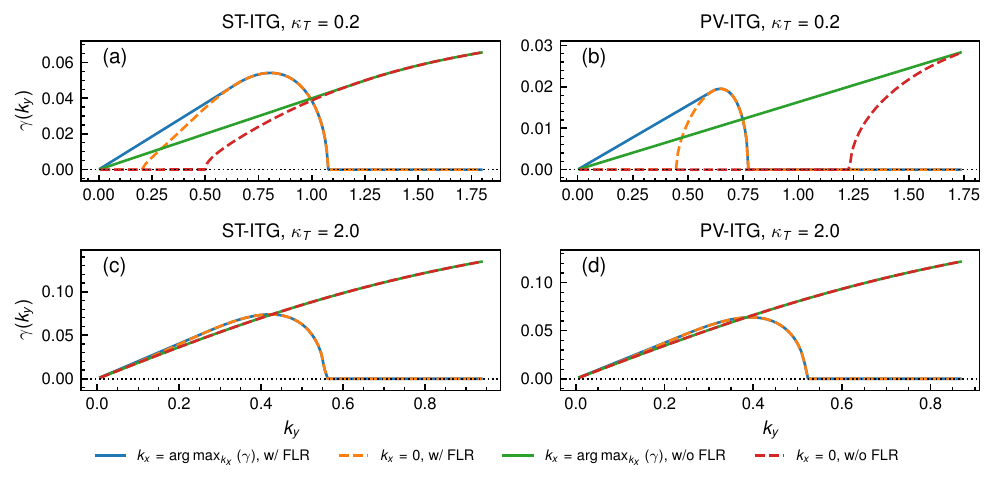}
\par\end{centering}
\caption{The linear growth rate, $\gamma(k_{y})$, for $\nu=0$ and $H=0$
for $\kappa_{T}=0.2$ and $\kappa_{T}=2.0$, for both models. Here
w/ and w/o mean with and without respectively. For all four cases,
the growth doesn't reduce to zero for large $k_{y}$ in the absence
of FLR terms (green and red). For $\kappa_{T}=0.2$ {[}(a) and (b){]},
$k_{x}=0$ is not the mode that maximizes $\gamma(k_{y})$ (with FLR)
for $k<k_{y,\mathrm{max}}$ unlike $\kappa_{T}=2.0$ {[}(c) and (d){]}
where there isn't any significant difference in $\gamma(k_{y})$ with
and without FLR terms. In addition, the maximum of the ST-ITG growth
rate is higher than that of PV-ITG for $\kappa_{T}=0.2$ but the maximum
growth rates of both models are comparable for $\kappa_{T}=2.0$.\label{fig:gam_vs_ky}}
\end{figure*}

To obtain $\gamma(k_{y})$, we pick the $k_{x}$ at each $k_{y}$
that maximizes $\gamma(k_{x},k_{y})$ at that $k_{y}$. A plot of
$\gamma(k_{y})$ in the dissipationless limit is shown in figure~\ref{fig:gam_vs_ky}.
For $\gamma(k_{y})$ with the FLR terms, the maximum is at $k_{y}=0.69$
for ST-ITG and $k_{y}=0.58$ for PV-ITG, which is roughly double of
what is reported using gyrokinetic codes~\citep{Dimits_2000}. For
$\kappa_{T}=2.0$, the maximum is at $k_{y}=0.42$ for ST-ITG and
$k_{y}=0.39$ for PV-ITG. Note also that the poloidal wavenumber that
maximizes the linear growth rate decreases as $\kappa_{T}$ increases.
Without the FLR terms (green and red curves), the growth rate keeps
increasing as a function of $k_{y}$, and is brought down to negative
values at $k_{y}>1$ only by the viscous dissipation. In contrast,
with the FLR terms, (blue and orange curves), the growth rate becomes
negative at high $k_{y}$ by itself because of these terms. This is
true for both the ST-ITG system and the PV-ITG system regardless of
the value of $\kappa_{T}$.

We also observe that there is an effect of whether we use $k_{x}=0$
or $k_{x}=\arg\max_{k_{x}}[\gamma(k_{x},k_{y})]$ to obtain $\gamma(k_{y})$.
One generally sets $k_{x}=0$ assuming that it maximizes the growth
rate. But as we observe, this is not exactly true at all scales and
is affected by the FLR terms as well as the temperature gradient.
We observe that for both systems the $k_{x}$ that maximizes the growth
rate at low $k_{y}$ is nonzero and the difference in $k_{x}$ vanishes
beyond the maximum. The maximization at finite $k_{x}\neq0$ is due
to the $(\tau+k^{2}_{x}+k^{2}_{y})$ from $(\tau\widetilde{\phi}-\nabla^{2}\phi)$
not canceling out, leading to a $k_{x}$ optimization of the growth
rate for each $k_{y}$, which is more important for small $k^{2}$.

\subsection{Mixing Length Estimate}

\begin{figure*}
\begin{centering}
\includegraphics{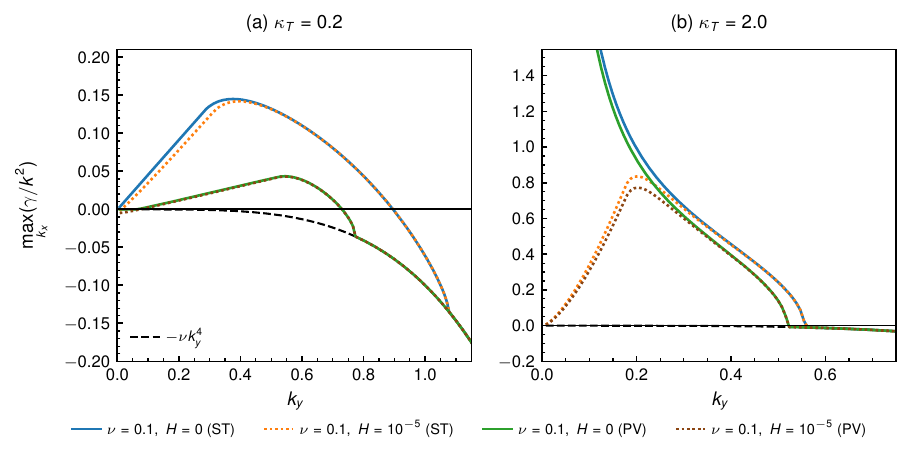}
\par\end{centering}
\caption{$\mathrm{max}_{k_{x}}(\gamma/k^{2})$ vs. $k_{y}$ plots for temperature
gradients below and above the nonlinear threshold for the ST-ITG and
PV-ITG system. (a) $\kappa_{T}=0.2$: the existence of a finite $(\gamma/k^{2})_{\mathrm{max}}$
doesn't depend on hypoviscous dissipation and the magnitude of $\mathrm{max}(\gamma/k^{2})$
of the ST-ITG model is higher than that of the PV-ITG model. (b) $\kappa_{T}=2.0$:
the system requires a hypoviscosity to have a finite $D_{\mathrm{turb}}$.
The hypoviscosity creates a negative growth rate at small $k_{y}$
mimicking the effect of large-scale damping resulting in a positively
sloped $\mathrm{max}_{k_{x}}(\gamma/k^{2})$ at low $k_{y}$ and thereby
a finite $D_{\mathrm{turb}}=\max(\gamma/k^{2})$.\label{fig:Dturbky}}
\end{figure*}

We generally study plasma turbulence, to estimate the transport it
generates. The simplest estimate of turbulent transport relies on
using a diffusive ansatz, and approximating the diffusion coefficient
using mixing-length arguments~\citep{Kadomstev_1965}, through resonance
broadening theory~\citep{Dupree_1966,Dupree_1967,Weinstock_1969,Weinstock_1970},
which gives $D_{\mathrm{turb}}\sim(\gamma/k^{2})_{\mathrm{max}}$.
To plot $\gamma/k^{2}$ as a function of $k_{y}$, we maximize $\gamma/k^{2}$
with respect to $k_{x}$ at each $k_{y}$ (shown in figure~\ref{fig:Dturbky})
and the maximum of this curve gives the mixing-length estimate for
the diffusion coefficient, $D_{\mathrm{turb}}$. For $\kappa_{T}=0.2$
{[}figure~\ref{fig:Dturbky}(a){]}, both the ST-ITG and PV-ITG systems
have a finite $D_{\mathrm{turb}}$ with (solid) and without hypoviscosity
(dotted), with the ST-ITG model {[}figure~\ref{fig:Dturbky}(a) solid
blue and dotted orange curves{]} having a higher $D_{\mathrm{turb}}$
than the PV-ITG one {[}figure~\ref{fig:Dturbky}(a) solid green and
dotted red curves{]}. Furthermore, ST-ITG model's $\max_{k_{x}}(\gamma/k^{2})$
curve peaks at a lower $k$ compared to that of the PV-ITG model suggesting
larger scale structures, which explains the higher transport coefficients.
Both models after $k=1.05$ (ST-ITG) and $k=0.75$ (PV-ITG) simply
trace the $-\nu k^{4}_{y}$ curve as the hyperviscosity dominates.

Moving on to the $\kappa_{T}=2.0$ case {[}figure~\ref{fig:Dturbky}(b){]},
we find that hypoviscosity is essentially necessary for the existence
of a finite $D_{\mathrm{turb}}$. When the hypoviscosity is zero for
the ST-ITG {[}figure~\ref{fig:Dturbky}(b) solid blue curve{]} and
PV-ITG {[}figure~\ref{fig:Dturbky}(b) solid green curve{]} models,
the $\max_{k_{x}}(\gamma/k^{2})$ curve diverges as $k_{y}\to0$.
This means that if the nonlinear dynamics follows the mixing-length
estimate, the system would generate larger and larger scales, and
never saturate. Whereas when we add finite hypoviscosity ($H=10^{-5}$),
the ST-ITG curve {[}figure~\ref{fig:Dturbky}(b) dotted orange curve{]}
and the PV-ITG curve {[}figure~\ref{fig:Dturbky}(b) dotted red curve{]}
can be made to decrease to zero as $k_{y}\to0$, with a well defined
peak, resulting in a finite $D_{\mathrm{turb}}.$ This clearly demonstrates
the necessity for large scale hypoviscosity for the turbulent saturation
in both the models before performing any nonlinear simulations. The
hypoviscosity provides effective large-scale damping that may mimic
kinetic damping missing in our fluid model, although its particular
form should not be interpreted as a quantitative model of kinetic
damping. We find that, for these parameters the $D_{\mathrm{turb}}$
for ST-ITG is higher than that of PV-ITG, even though the $\max_{k_{x}}(\gamma/k^{2})$
curves of both the models with a given finite hypoviscosity peak around
the same $k$ unlike the $\kappa_{T}=0.2$ case. Again, both curves
simply trace the $-\nu k^{4}_{y}$ curve after $k_{y}\approx0.5$.

\subsection{Instability Threshold}

Another useful criterion one can obtain from a linear analysis is
the threshold of instability. For example, the ST-ITG system becomes
unstable if
\begin{align}
(\kappa_{n}+\kappa_{T})^{2}k^{4}-2(\kappa_{n}+\kappa_{T})(\kappa_{n}+2\kappa_{B})k^{2}+\kappa^{2}_{n}-4\kappa_{B}\tau(\kappa_{n}+\kappa_{T}) & <0\ \text{for some}\ k.
\end{align}

Hence, the instability condition for $\kappa_{n}>-2\kappa_{B}$ is
$\kappa_{T}>-\kappa_{n}$ and for $\kappa_{n}\le-2\kappa_{B}$ it
is $\kappa_{T}>\kappa^{2}_{n}/4\kappa_{B}\tau-\kappa_{n}$. Therefore,
the system is always unstable in the $\kappa_{n}>0$, $\kappa_{T}>0$
domain. On the other hand, the condition for instability in the dissipationless
limit for the PV-ITG system is
\begin{equation}
[(\kappa_{n}+\kappa_{T})^{2}+4\kappa_{B}\Gamma(\kappa_{n}+\kappa_{T})]k^{4}-2(\kappa_{n}+\kappa_{T})(\kappa_{n}+[2+(1-2\tau)\Gamma]\kappa_{B})k^{2}+(\Gamma\kappa_{B}+\kappa_{n})^{2}-4\kappa_{B}\tau(\kappa_{n}+\kappa_{T})<0\ \text{for some}\ k.
\end{equation}

Unlike the ST-ITG system, the PV-ITG system has a $\kappa_{T}/\kappa_{n}=2/3$
threshold line in the positive gradient domain. For toroidal ITG,
this linear instability threshold arises from the $\Gamma(\kappa_{n}+\kappa_{T})\partial_{y}\nabla^{2}\phi$
term in equation~(\ref{eq:itg2d_pv-1}), which is an $\mathcal{O}(\epsilon)$
FLR term that is generally neglected. Note that while neglecting the
FLR terms in both equations yields the same instability threshold,
doing so would be inconsistent with the ordering of the potential
equation. Therefore, to observe the $\kappa_{T}/\kappa_{n}=2/3$ instability
threshold while having a consistently ordered potential equation,
we need to retain FLR terms in the pressure equation. Strictly from
the standpoint of preserving the $2/3$ threshold, the $-\Gamma\kappa_{B}\partial_{y}P$
term in the pressure equation is unnecessary as long as the FLR term
is retained. However, if one wishes to relax the large aspect ratio
ordering (move to higher $\kappa_{B}$) while preserving the $2/3$
threshold, the curvature term in the pressure equation should be retained
like in the PV-ITG system.

\subsection{The choice of dissipation parameters}

The linear system also allows us to choose simulation parameters so
that it damps large- and small-scale modes, and provides a wavenumber
range that is large enough to resolve both the large scale modes,
such as zonal flows or streamers and the turbulent cascade eventually
terminating at dissipative scales. Let $k_{y,\mathrm{max}}$ be the
wavenumber $k_{y}$ at which the linear growth rate is maximized (i.e.
$\gamma=\gamma_{\mathrm{max}}$), and $k_{\mathrm{cutoff}}$ be the
wavenumber where $\gamma$ drops to zero, marking the end of the linear
energy injection range. To ensure that the simulation adequately resolves
the forward cascade, we set our maximum resolved wavenumber to at
least $3k_{\mathrm{cutoff}}$ such that it necessarily includes a
range of $k$ after injection ends and dissipation begins. However
note that the inaccuracy of the simple fluid FLR nonlinearities that
we use for high $k$ limits this range in that, we can not really
go to normalized wavenumbers that are much larger than $1$. Note
that this is in stark contrast to the Hasegawa-Wakatani model, where
one recovers the two-dimensional Navier-Stokes system for $k\rho_{s}\gg1$,
which is nicely well behaved, even though not particularly accurate
for plasma turbulence.

Going back to a purely linear perspective, the viscous dissipation
coefficient $\nu$ must satisfy two competing criteria. First, to
avoid artificially suppressing linear physics, the viscous damping
rate at the injection scale must be much smaller than the instability
drive: $\nu k^{6}_{\mathrm{y,\mathrm{max}}}\ll\gamma_{\mathrm{max}}$
(e.g. something like $100$ times in practice seems to be fine). Second,
the damping rate at the grid scale (which in the limiting case is
$k_{\mathrm{grid}}\approx3k_{\mathrm{cutoff}}$) must remain finite
to ensure that the energy is dissipated. The most restrictive bounds
on $\nu$ occur at the lowest temperature gradient that we use ($\kappa_{T}=0.2$).
For this temperature gradient, in the dissipationless limit, $\gamma_{\mathrm{max}}=0.054$,
$k_{y,\mathrm{max}}=0.805$, and $k_{\mathrm{cutoff}}=1.078$. Choosing
$\nu\sim10^{-3}$ satisfies $\nu k^{6}_{y,\mathrm{max}}\ll\gamma_{\mathrm{max}}$
while keeping grid-scale dissipation finite ($\nu k^{6}_{\mathrm{grid}}\approx1.14\gamma_{\mathrm{max}}$).
However, we end up choosing an extremely high hyperviscosity coefficient
as discussed in later sections, $\nu=0.1$ ($\nu k^{6}_{y,\mathrm{max}}=0.504\gamma^{\nu=0}_{\mathrm{max}}$),
that distorts $\gamma(k_{y})$, by reducing its magnitude and increasing
the damping on medium wavenumbers. This is necessary to damp the FLR
nonlinearities at high $k$, where the simple fluid approximation
that we use appears to be inaccurate, resulting in spurious small-scale
numerical instabilities.

\section{Nonlinear simulations\label{sec:Nonlinear-simulations}}

A pseudo-spectral code, based on the code used by reference~\citealp{Gurcan_2022}
for direct numerical simulations of the Hasegawa-Wakatani System,
was developed and used to simulate the 2D ITG system. The initial
conditions for the Fourier amplitudes of the perturbed fields were
taken to be small amplitude Gaussian distributions, $10^{-6}\exp(-k^{2}/100)$,
with randomly assigned phases. For the numerical simulations shown
below, all the physical parameters are fixed and the only parameter
that is varied is the non-dimensionalized background temperature gradient,
$\kappa_{T}$.

The simulation parameters, inspired by the CYCLONE base case (CBC)~\citep{Dimits_2000},
are listed in table~\ref{tab:sim_params}. The parameter that we
manually choose is $\kappa_{B}$ so that we enforce large-aspect-ratio
ordering. We choose a round value of $\kappa_{B}=1/50$ corresponding
to $\rho_{*}\coloneqq\rho_{s}/a=1/36$ for an aspect ratio $a/R=0.36$
and ion-to-electron temperature ratio $\tau=1$. Consequently, our
dimensionless background gradients are related to the standard non-dimensionalized
background gradients $(R/L_{n/T})$~\citep{Dimits_2000} as $\kappa_{n,T}=(\kappa_{B}/2)R/L_{n/T}=0.01R/L_{n/T}$.

Typical ITG turbulence correlation lengths from experimental measurements
are $l_{c}=5-10\rho_{i}$~\citep{Rhodes_2002} and we need a box
size to be much larger than the correlation length. Benchmark simulations
show that there was very little change in the simulation runs as the
perpendicular size $125\rho_{i}$ was doubled or halved. This spectral
requirement implies a safe minimum physical domain size of $L_{\mathrm{phys}}\approx63\rho_{i}$.
Our simulation uses a physical box size of $L_{\mathrm{phys}}=64\pi\rho_{i}\approx200\rho_{i}$,
which exceeds these standard requirements and prevents artificial
self-correlation of the turbulent structures. The box size also needs
to be small compared to the device size ($\sim a$, the minor radius)
for the `local' gradient-driven assumption to hold. The ion gyroradius
is of the order of a mm, therefore $L_{\mathrm{phys}}$ is of the
order $0.2\mathrm{m}$. Consequently, the box size is lower than the
minor radius ($a\sim1\mathrm{m}$) of a typical fusion device. In
other words, while it doesn't make sense to directly model the fluctuations
in a tokamak using our simple 2D fluid model, it passes the sanity
check requirements to be an ITG model for fusion plasmas.

\begin{figure*}
\centering{}\includegraphics{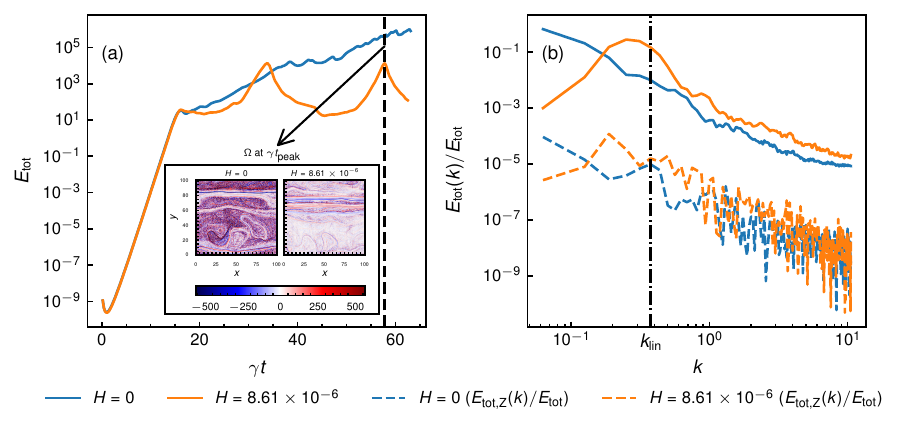}\caption{Comparison between simulations with hypoviscosity (orange) and without
hypoviscosity (blue) where dissipation is of the form: $0.1\nabla^{2}Y-H\nabla^{-4}\widetilde{Y}$
with $L=32\pi$ and $N_{x},N_{y}=340$. (a) Time evolution of the
total energy of the system with hypoviscosity (orange) and without
hypoviscosity (blue). The system without hypoviscosity continues to
grow and the simulation diverges. The time instant ($\gamma t_{\mathrm{peak}}=57.81$)
was identified by computing when the total energy is maximum for the
simulation with the hypoviscosity as shown by the vertical dash line.
The inset shows the vorticity of both systems with and without hypoviscosity
at $t_{\mathrm{peak}}$. (b) Total energy spectra during a streamer
dominant state are shown. For low-$k$, the curve without hypoviscosity
continues to increase as $k$ decreases whereas the curve of the simulation
with hypoviscosity decreases.\label{fig:hypo_comp}}
\end{figure*}

We use the standard 2/3 dealiasing so that the maximum valid wavenumber
is $(2/3)\pi/\Delta x=k_{\mathrm{grid}}$ ($\pi/\Delta x$ is the
Nyquist mode). The minimum resolution required for a simulation with
a box size $L$ to resolve up to at least $k_{\mathrm{grid}}$ is
$N_{\mathrm{min}}=L/\Delta x=3k_{\mathrm{grid}}L/(2\pi)$. For $L=64\pi$
and $k_{\mathrm{grid}}\approx3k_{\mathrm{cutoff}}$, $N_{\mathrm{min}}\approx310$,
which gives a padded minimum of $N_{\mathrm{p,min}}\approx465$. Rounding
off the padded minimum to the next power of $2$, to increase the
speed of the fast Fourier transforms, we get $N_{\mathrm{p,min}}=512$.
However, the initial simulation {[}figure~\ref{fig:hypo_comp}{]}
to motivate the use of hypoviscosity and hyperviscosity in place of
Laplacian viscosity uses a limited box size of $L=32\pi$ and resolution
$N_{x},N_{y}=340$ ($N_{px},N_{py}=512$) unlike the simulations that
follow, for model comparisons, which use $L$ and $N_{x},N_{y}$ given
in table~\ref{tab:sim_params}. In practice, any hyperviscosity coefficient
$\nu$ below $0.01$ leads to numerical instabilities for PV-ITG and
a $\nu$ below $0.1$ results in a lower zonal kinetic energy fraction
for ST-ITG (see section~\ref{sec:Viscous-dissipation}). Therefore
$\nu=0.1$, which is quite high, is chosen as the coefficient of hyperviscosity
for the simulations. Such a high hyperviscosity kills off most of
the $k>1$ dynamics, which is consistent with the low wavenumber assumption
used to derive the fluid model. Numerically, using a high hyperviscosity
coefficient makes the problem stiff and therefore we use a modified
(for stiff PDEs) fourth-order exponential time-differencing Runge-Kutta
(ETDRK4) scheme~\citep{Gurcan_etdrk4cp_github} based on reference~\citealp{Kassam_2005}.
The simulation scripts used in this study are openly available~\citep{Paramasivam_itg2d-flr-fluid_github}
and are archived on Zenodo~\citep{Paramasivam_itg2d-flr-fluid_2026}.

\begin{table}
\centering{}%
\begin{tabular}{|c|c|c|c|c|}
\hline 
Parameters & \multicolumn{2}{c|}{ST-ITG} & \multicolumn{2}{c|}{PV-ITG}\tabularnewline
\hline 
$N_{px},N_{py}$ & \multicolumn{2}{c|}{$1024$} & \multicolumn{2}{c|}{$1024$}\tabularnewline
\hline 
$N_{x},N_{y}$ & \multicolumn{2}{c|}{$682$} & \multicolumn{2}{c|}{$682$}\tabularnewline
\hline 
$L_{x},L_{y}$ & \multicolumn{2}{c|}{$64\pi$} & \multicolumn{2}{c|}{$64\pi$}\tabularnewline
\hline 
$\tau$ & \multicolumn{2}{c|}{$1$} & \multicolumn{2}{c|}{$1$}\tabularnewline
\hline 
$\Gamma$ & \multicolumn{2}{c|}{$5/3$} & \multicolumn{2}{c|}{$5/3$}\tabularnewline
\hline 
$\kappa_{B}$ & \multicolumn{2}{c|}{$0.02$} & \multicolumn{2}{c|}{$0.02$}\tabularnewline
\hline 
$\kappa_{n}$ & \multicolumn{2}{c|}{$0.2$} & \multicolumn{2}{c|}{$0.2$}\tabularnewline
\hline 
$\kappa_{T}$ & $0.2$ & $2.0$ & $0.2$ & $2.0$\tabularnewline
\hline 
$D$ & $0.1$ & $0.1$ & $0.1$ & $0.1$\tabularnewline
\hline 
$\nu$ & $0.1$ & $0.1$ & $0.1$ & $0.1$\tabularnewline
\hline 
$H$ & $1\times10^{-5}$ & $4\times10^{-5}$ & $1\times10^{-5}$ & $4\times10^{-5}$\tabularnewline
\hline 
\end{tabular}\caption{Simulation parameters: $N_{px,y}$ is the padded resolution and $N_{x,y}=2\lfloor N_{px,y}/3\rfloor$
corresponding to a maximum wavenumber of $k_{x,y,\mathrm{max}}=2\pi N_{x,y}/L_{x,y}$.
$D$ and $\nu$ are the viscosity coefficients of $P$ and $\tau\widetilde{\phi}-\nabla^{2}\phi$
respectively. $H$ is the hypoviscous coefficient and is larger for
$\kappa_{T}=2.0$ to strongly damp large scales.\label{tab:sim_params}}
\end{table}

For the simulations with half the box size and resolution of the values
given in table~\ref{tab:sim_params}, and a Laplacian viscosity coefficient
of $\nu_{\mathrm{2}}=0.1$, we see that for low $\kappa_{T}$, in
the zonal-flow-dominated regime, the system exhibits a saturated steady
state, in the sense that there are no big bursts in any of the conserved
quantities or the heat flux. However, beyond the nonlinear threshold
($\kappa_{T,\mathrm{nl}}$), where zonal flows are minimal, the system
forms box-sized streamers and hence does not reach steady state. We
fix this by adding a hypoviscosity term with a coefficient big enough
to damp large-scale non-zonal modes partly to mimic large-scale kinetic
damping~\citep{Nordman_1989,Nordman_1990}. This additional large-scale
damping is necessary for the 2D ITG system in contrast with the Hasegawa-Wakatani
system that has a term of the form $-C\phi_{k}/k^{2}$ in the vorticity
equation, where $C$ is the adiabaticity constant, arising from the
divergence of the parallel current, that damps large scale modes,
naturally.

First, we show the results of simulations with and without the hypoviscosity
term ($H\nabla^{-4}Y$, $H=8.6\times10^{-6}$) for $\kappa_{T}=2.0$
to illustrate its importance for steady state saturation {[}figure~\ref{fig:hypo_comp}(a){]}.
We observe that when we do not use hypoviscosity (blue), the simulation
simply blows up, which is somewhat expected of 2D turbulence, but
faster because of the additional nonlinear terms. However, if hypoviscosity
is used (orange), it returns from the streamer dominated state to
a well behaved turbulent state before it surges again ($\gamma t_{\mathrm{peak}}=57.8$),
when streamers become dominant again, and then the turbulence recovers
once again to steady levels. Furthermore, we can observe the effect
of the hypoviscosity on the spectrum in figure~\ref{fig:hypo_comp}(b)
which shows the instantaneous total energy spectra {[}equation~(\ref{eq:energy_spectrum}){]}
of both the simulations at $\gamma t_{\mathrm{peak}}$. The hypoviscous
term damps the large scales resulting in a positively sloped total
energy spectrum at low-$k$. This makes it so that even when a streamer
induces a surge in the energy and heat flux, energy is not accumulated
in the largest scales allowing the system to recover. This results
in a bursty evolution of the system (orange) in contrast to the system
without hypoviscosity (blue), where the simulation would simply blow
up. From this, we conclude that some sort of hypoviscosity is necessary
for the saturation of the simulation in the high-transport turbulent
regime.

Two snapshots from the two simulations are shown at the second peak
of the finite $H$ simulation, in the inset of figure~\ref{fig:hypo_comp}(a).
We observe a large scale, radially extended nonlinear structure whose
radial extent is comparable to the box size. These streamers ($k_{x}\ll1$
modes) are associated with enhanced transport exceeding the levels
predicted from turbulence saturation estimates ($\sim(\gamma/k^{2})_{\mathrm{max}}$)~\citep{Drake_1988}.
They are equivalent to convective cells, in Rayleigh-Bénard turbulence
and dominate the transport whenever they form, since they can only
appear when the zonal flows are absent. We also observe a lot of small-scale
noise even in the finite $H$ simulation when it approaches its peak.
This is due to the fact that the Laplacian viscosity is not strong
enough to damp out the small scales completely in the presence of
the additional nonlinear terms. We see this in the instantaneous total
energy spectra of both the simulations at $\gamma t_{\mathrm{peak}}$
in figure~\ref{fig:hypo_comp}(b) where the spectrum is not sufficiently
damped at large wavenumbers. Hence, we require hyperviscosity instead
of Laplacian viscosity to avoid these extreme bursts.

\section{Role of FLR nonlinearities\label{sec:Role-of-flr-nl}}

As noted in the introduction, the diamagnetic stress can be significant
and models that omit it may not be able to capture the zonal flow
physics of the ITG system even qualitatively. For example, a $\kappa_{T}=2.0$
simulation, which should yield a high-transport state, is zonal flow
dominant without the diamagnetic nonlinearity. However, the introduction
of an additional nonlinearity in a two-dimensional turbulence model
brings with it new problems regarding saturation and spectral energy
transfer. This is further exacerbated by the fact that the form of
the additional nonlinearity ruins the traditional 2D turbulence behavior
of the otherwise advective system. In particular the system no longer
reduces to a simple 2D Navier-Stokes equation in small scales because
of the scale dependence of the newly introduced FLR nonlinearities.
Therefore, to characterize its behavior, the evolution, the spectra,
and the transfers of conserved quantities are studied in this section
to identify the contribution of the diamagnetic nonlinearity.

To explicitly identify the contributions of the Reynolds and diamagnetic
stresses to the zonal flow dynamics, we take the zonal average of
the potential equation, which is the same for both models, and integrate
once with respect to $x$, obtaining the zonal flow evolution equation,
\begin{align*}
\partial_{t}\partial_{x}\overline{\phi} & =-\partial_{x}R_{\phi}-\partial_{x}R_{d}+\nu\partial^{6}_{x}\partial_{x}\overline{\phi}\,\text{,}
\end{align*}

where $\partial_{x}\overline{\phi}$ is the zonal velocity, $R_{\phi}=-\overline{\partial_{x}\phi\partial_{y}\phi}$
is the ($\mathrm{E}\times\mathrm{B}$) Reynolds stress, and $R_{d}=-\overline{\partial_{x}\phi\partial_{y}P}$
is the diamagnetic stress. Here, the overbar $\overline{\cdots}$
represents the zonal average. The diamagnetic stress arises from the
diamagnetic nonlinearity in the potential equation, and its role in
zonal flow destabilization at the transition to the high-transport
regime has been discussed in reference~\citealp{Ivanov_2020}. The
hyperviscous term represents zonal flow damping, while the hypoviscosity
does not contribute because it acts only on nonzonal modes. Although
the zonal flow equation has the same form for both models, differences
in their pressure dynamics and linear properties can modify the resulting
turbulent stresses.

\subsection{Time evolution\label{subsec:Time-evolution}}

The zonal kinetic energy fraction, $\Xi\coloneqq E_{\mathrm{kin,ZF}}/E_{\mathrm{kin}}$,
is commonly used to characterize the zonal flow state of drift-wave
turbulence~\citep{Numata_2007,Guillon_2025}. For the ITG case, a
similar quantity can be used to characterize the Dimits shift. In
gyrokinetics, this takes the form of the zonal free energy fraction~\citep{Marquant_2026},
whereas in our 2D fluid model the zonal kinetic energy fraction, even
though kinetic energy is not conserved, seems to be the most convenient
choice. In this perspective, if the temperature gradient is far below
the critical value for the nonlinear transition, but above the linear
instability threshold, $\Xi$ is close to $1$. In contrast, if the
temperature gradient is well above this value, the $\Xi$ is close
to zero. In this sense, $\Xi$ also defines if the system is close
to, or far from marginality in a qualitative sense. It seems however
that, unlike the drift-wave case, for ITG, the transition is less
abrupt and intermediate values of the zonal kinetic energy fraction
are found around the critical value. In this paper, we stay away from
such values of the temperature gradient, which may lie in a region
corresponding to a hysteresis loop, in order to maintain two identifiably
distinct regimes.

\begin{figure*}
\centering{}\includegraphics{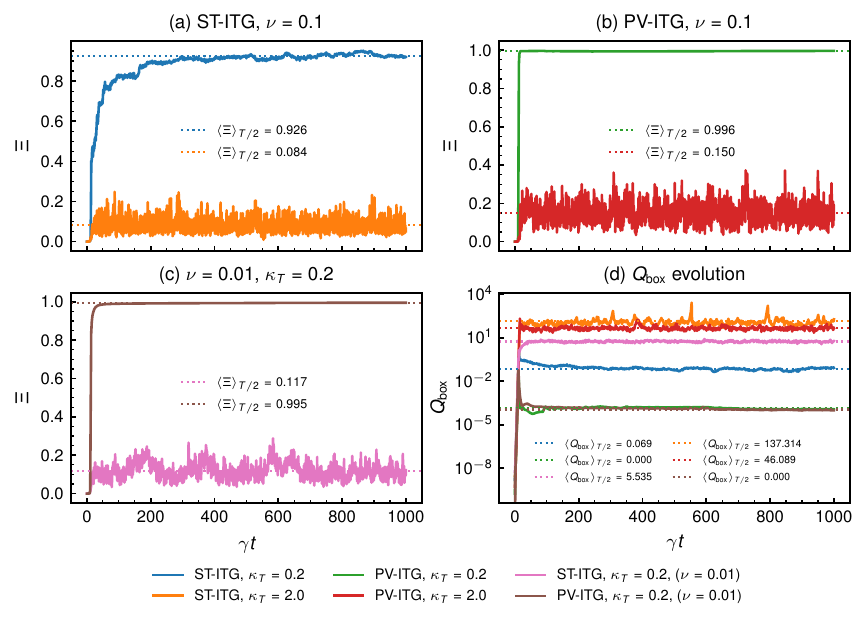}\caption{(a) Zonal kinetic energy fraction, $\Xi$, vs. $\gamma t$ for ST-ITG:
For $\kappa_{T}=0.2$ (blue), $\langle\Xi\rangle_{T/2}$ is $0.926$
whereas for $\kappa_{T}=2.0$ (orange), it is only $0.084$ and the
fluctuations in the evolution are comparatively large. (b) $\Xi$
vs. $\gamma t$, PV-ITG: For $\kappa_{T}=0.2$ (green), $\langle\Xi\rangle_{T/2}$
is $0.996$, which is slightly higher than the ST-ITG model's fraction,
and it reaches this value quicker. For $\kappa_{T}=2.0$ (red), $\langle\Xi\rangle_{T/2}$
is $0.150$, which is roughly double that of the ST-ITG simulation.
(c) $\Xi$ vs. $\gamma t$, $\nu=0.01$: For ST-ITG, $\langle\Xi\rangle_{T/2}$
reduces to $0.117$ and for PV-ITG, it remains the same at around
$0.995$. However, the PV-ITG system takes slightly longer to saturate
for $\nu=0.01$ compared to $\nu=0.1$. (d) Box-averaged heat flux,
$Q_{\mathrm{box}}$, vs. $\gamma t$. ST-ITG: For $\kappa_{T}=0.2$
(blue), $\langle Q_{\mathrm{box}}\rangle_{T/2}$ is $0.069$ and increases
to $137.314$ for $\kappa_{T}=2.0$ (orange). PV-ITG: For $\kappa_{T}=0.2$
(green), $\langle Q_{\mathrm{box}}\rangle_{T/2}$ is zero and increases
to $46.089$ for $\kappa_{T}=2.0$ (red), which is roughly one-third
of that of the ST-ITG model. $\nu=0.01$ and $\kappa_{T}=0.2$: The
$\langle Q_{\mathrm{box}}\rangle_{T/2}$ of the ST-ITG (cyan) is $5.535$
which is higher than that of the $\nu=0.1$ simulation. However, the
$\langle Q_{\mathrm{box}}\rangle_{T/2}$ of PV-ITG (brown) is zero
just like $\nu=0.1$ simulation of the same.\label{fig:zonal_KE_frac_Q_vs_t_comp_kapt_PV}}
\end{figure*}

The evolution of the zonal kinetic energy fraction of the ST-ITG and
PV-ITG models for $\kappa_{T}=0.2$ and $\kappa_{T}=2.0$ is shown
in figure~\ref{fig:zonal_KE_frac_Q_vs_t_comp_kapt_PV}(a) and~(b)
respectively. The evolution of box-averaged heat fluxes, $Q_{\mathrm{box}}=\langle-\partial_{y}\phi P\rangle$,
is shown in figure~\ref{fig:zonal_KE_frac_Q_vs_t_comp_kapt_PV}(d)
for the corresponding simulations. Note that for adiabatic electrons,
$T=P-\tau\widetilde{\phi}$ and hence $\overline{\partial_{y}\phi T}=\overline{\partial_{y}\phi P}$.
In order to achieve two clearly distinct regimes as characterized
by the zonal kinetic energy fraction, we consider $\kappa_{T}=0.2$
and $\kappa_{T}=2.0$ {[}blue and orange curves in figure~\ref{fig:zonal_KE_frac_Q_vs_t_comp_kapt_PV}(a)
respectively{]}. The mean of the zonal kinetic energy fraction over
the second half of the ST-ITG simulation, $\langle\Xi\rangle_{T/2}$,
for $\kappa_{T}=0.2$ is $0.926$ while it drops to $0.084$ for $\kappa_{T}=2.0$.
We also see a corresponding change in the mean of the box-averaged
heat flux, $\langle Q_{\mathrm{box}}\rangle_{T/2}$, which increases
by more than three orders of magnitude from a meager $0.069$ for
$\kappa_{T}=0.2$ to $137.314$ for $\kappa_{T}=2.0$. Furthermore,
we observe that both the zonal kinetic energy fraction and the heat
flux fluctuate much more in the turbulent regime, with $\kappa_{T}=2.0$,
than in the zonal-flow-dominated regime, with $\kappa_{T}=0.2$.

We see that the PV-ITG model replicates the same trend for $\kappa_{T}=0.2$
and $\kappa_{T}=2.0$ {[}green and red curves in figure~\ref{fig:zonal_KE_frac_Q_vs_t_comp_kapt_PV}(b)
respectively{]}, with a mean zonal kinetic energy fraction of $0.996$
for $\kappa_{T}=0.2$ and $0.150$ for $\kappa_{T}=2.0$. In both
cases, the PV-ITG system generates more zonal flows, which in turn
suppresses the flux more efficiently than the ST-ITG system. The corresponding
heat fluxes are shown in figure~\ref{fig:zonal_KE_frac_Q_vs_t_comp_kapt_PV}(d)
for $\kappa_{T}=0.2$ (green) and $\kappa_{T}=2.0$ (red). $\langle Q_{\mathrm{box}}\rangle_{T/2}$
increases from zero (up to three decimal digits) for $\kappa_{T}=0.2$
to $46.089$ for $\kappa_{T}=2.0$. Note that $\langle Q_{\mathrm{box}}\rangle_{T/2}$
of $\kappa_{T}=2.0$ for PV-ITG is roughly one-third the value of
that of the ST-ITG simulation for $\kappa_{T}=2.0$. Finally, we observe
that both the zonal kinetic energy fraction and the heat flux have
larger fluctuations in the turbulent regime than in the zonal-flow-dominated
regime, just like those of ST-ITG. However, because the $\kappa_{T}=2.0$
case for PV-ITG has a higher zonal flow level, its heat flux fluctuates
less than that of ST-ITG, corresponding to a near-Gaussian kurtosis
of $4.718$ compared to that of $105.822$ for ST-ITG.

\begin{figure*}
\begin{centering}
\includegraphics{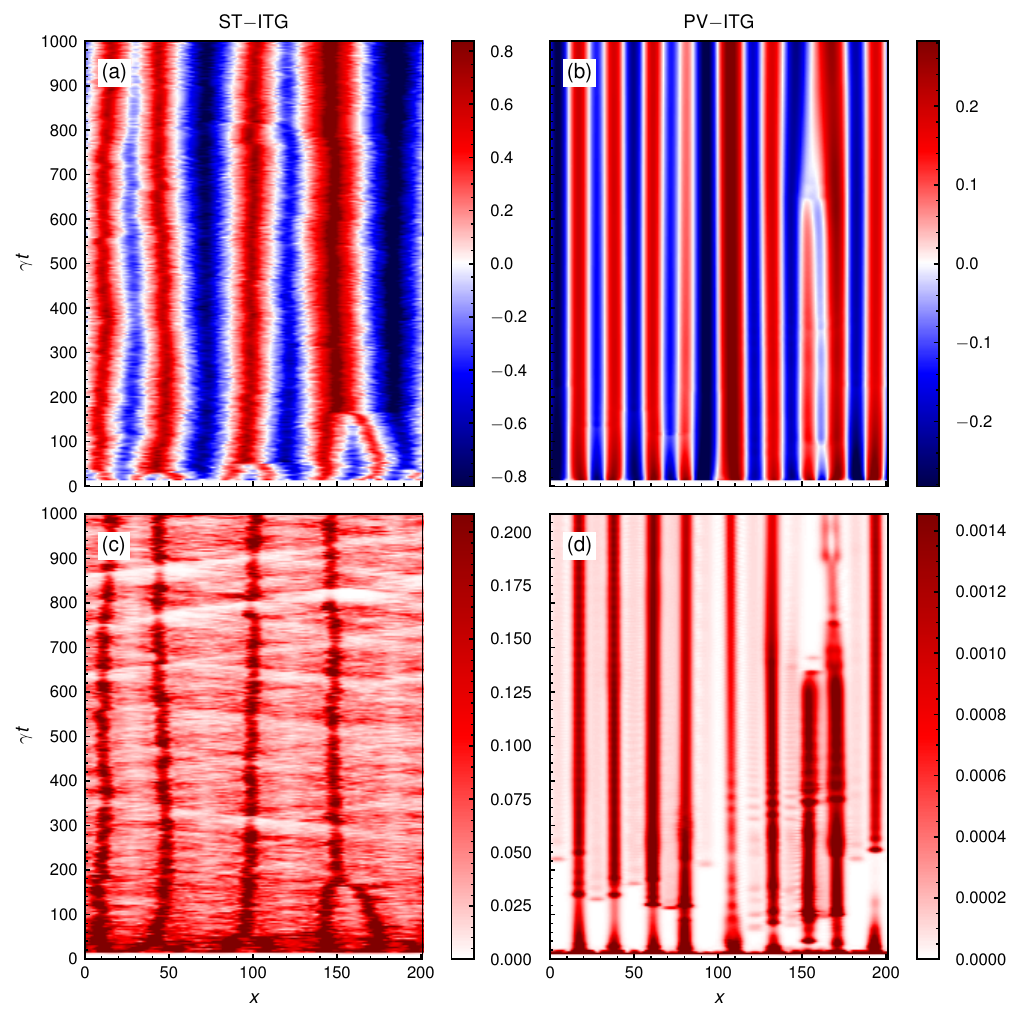}
\par\end{centering}
\caption{Spatiotemporal evolution of the zonal velocity (top row) and zonally
averaged turbulence intensity $\overline{\widetilde{\phi}^{2}}$ (bottom
row) for $\kappa_{T}=0.2$. (a) In the ST-ITG system, the zonal flows
oscillate and their maxima merge, mostly at $\gamma t<200$, resulting
in a reduction in their radial wavenumber. (b) Zonal flows remain
the same except for one major merger between $\gamma t=600$ and $\gamma t=700$.
(c,d) The turbulence intensity is substantially larger in the ST-ITG
system than in the PV-ITG system. Furthermore, the turbulence intensity
in the PV-ITG system is more strongly localized in radius. Note the
different color scales in (c) and (d).\label{fig:vbar_xt_comp_PV}}
\end{figure*}

The evolution of the zonal $\mathrm{E}\times\mathrm{B}$ velocity
of the ST-ITG system for $\kappa_{T}=0.2$ is shown in figure~\ref{fig:vbar_xt_comp_PV}(a).
We observe that the zonal flow maxima move a bit and merge especially
initially as the quasi-stationary pattern is formed. It appears that
it is the positive flows that merge, causing the zonal flow wavenumber
to decrease, accompanied by a jump in the zonal kinetic energy fraction.
For example, at $\gamma t\approx157$, when the zonal maxima merge
the zonal kinetic energy fraction jumps as seen in figure~\ref{fig:zonal_KE_frac_Q_vs_t_comp_kapt_PV}(a).
The time evolution of the zonal velocity is shown in figure~\ref{fig:vbar_xt_comp_PV}(b)
for the PV-ITG system with $\kappa_{T}=0.2$. Here the zonal flow
extrema rarely merge and as a result the zonal flow wavenumber stays
higher than that of the ST-ITG system. In contrast, there is no interesting
zonal flow dynamics for $\kappa_{T}=2.0$ (not shown), since turbulence
dominates and zonal flows present no temporal coherence.

The corresponding evolution of the zonally averaged turbulence intensity,
$\overline{\widetilde{\phi}^{2}}$, is shown in figures~\ref{fig:vbar_xt_comp_PV}(c)
and (d). The turbulence intensity is substantially larger in the ST-ITG
system than in the PV-ITG system, consistent with the higher maximum
linear growth rate of the ST-ITG system. In both systems, the maxima
of the turbulence intensity appear to follow the positive zonal flow
maxima, including their radial motion and mergers. Furthermore, the
turbulence intensity in the PV-ITG system is strongly localized in
narrow bands, whereas in the ST-ITG system it is more broadly distributed,
exhibits stronger temporal fluctuations, and displays features suggestive
of radially propagating structures.

It is important to note that the comparison between the ST-ITG and
PV-ITG systems at fixed parameters does not isolate the effect of
potential vorticity conservation on zonal flow formation and saturation.
As discussed in subsection~\ref{subsec:PV-ITG-model} and section~\ref{sec:Linear-Analysis},
restoring potential vorticity conservation requires retaining additional
linear and nonlinear terms in the pressure equation, which also modify
the linear instability characteristics and nonlinear interactions
of the system. In particular, the two models have different maximum
growth rates for $\kappa_{T}=0.2$. Therefore, the larger $\langle\Xi\rangle_{T/2}$
and lower $\langle Q_{\mathrm{box}}\rangle_{T/2}$ observed in the
PV-ITG simulations should be understood as properties of the complete
PV-ITG system rather than being attributed exclusively to potential
vorticity conservation. Nevertheless, for $\kappa_{T}=2.0$, where
the maximum linear growth rates of the two systems are comparable,
appreciable differences in their nonlinear saturated states persist.

\subsection{Spectra}

The instantaneous total energy spectrum is given by averaging the
modal energy, $E_{\boldsymbol{q}}(t)=(\tau\mathds{1}_{q_{y}\neq0}+q^{2})|\phi_{\boldsymbol{q}}|^{2}/2$,
over a shell of width $\Delta k$,
\begin{equation}
E_{\mathrm{tot}}(k,t)=\frac{1}{2\Delta k}\sum_{k-\Delta k/2<\left|\boldsymbol{q}\right|\le k+\Delta k/2}(\tau\mathds{1}_{q_{y}\neq0}+q^{2})|\phi_{\boldsymbol{q}}|^{2}\,\text{.}\label{eq:energy_spectrum}
\end{equation}

\begin{figure*}
\centering{}\includegraphics{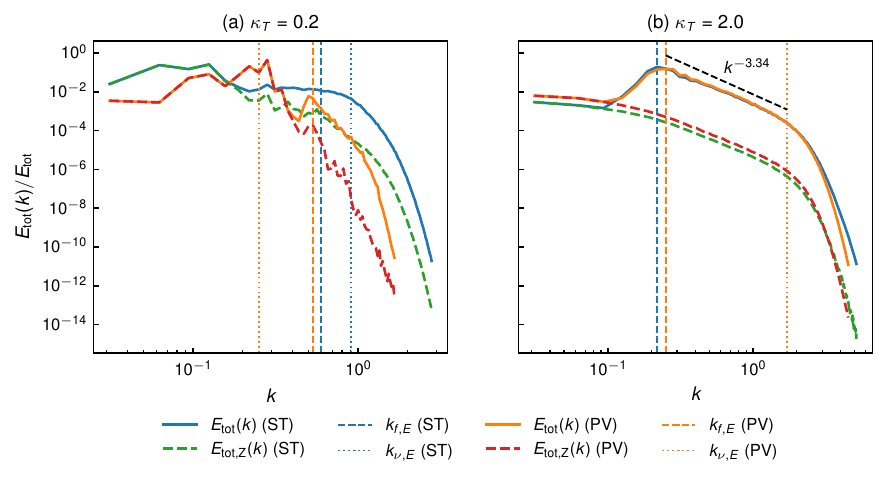}\caption{Normalized total energy spectrum, $E_{\mathrm{tot}}(k)/E_{\mathrm{tot}}$,
vs. the perpendicular wavenumber, $k$, for ST-ITG {[}solid blue and
dashed green curves{]} and PV-ITG {[}solid orange and dashed red curves{]}.
The dashed curves correspond to the zonal component of the spectra
normalized by the total energy. (a) For $\kappa_{T}=0.2$, the peaks
of the spectra are dominated by the zonal component of the spectrum.
Both $k_{\nu,E}$ are lower than $1.0$ with $k_{\nu,E}\text{(PV)}$
being lower than $k_{f,E}\text{(PV)}$. (b) For $\kappa_{T}=2.0$,
both the spectra are similar to one another, and have similar $k_{f,E}$
and $k_{\nu,E}$. The spectrum scales as $E_{\mathrm{tot}}(k)\sim k^{-3.24}$
for ST-ITG and $E_{\mathrm{tot}}(k)\sim k^{-3.34}$ for PV-ITG in
the domain $(k_{f,E},k_{\nu,E})$.\label{fig:Ek_comp_kapt_PV}}
\end{figure*}

Averaging the total energy spectra over time (from $\gamma t=500$
to $\gamma t=1000$) {[}$E_{\mathrm{tot}}(k)${]} and dividing by
their total energies {[}$E_{\mathrm{tot}}=\sum_{k}E_{\mathrm{tot}}(k)${]}
we get the normalized wavenumber spectra shown in figure~\ref{fig:Ek_comp_kapt_PV}
for both ST-ITG and PV-ITG models. For the $\kappa_{T}=0.2$ case
{[}figure~\ref{fig:Ek_comp_kapt_PV}(a){]}, the spectrum peaks, where
it is dominated by the zonal component (dashed green and dashed red
curves). The total energy spectrum of the ST-ITG model (solid blue
curve) is almost flat between the peak wavenumber where it is dominated
by the zonal component and the wavenumber where the spectrum falls
off due to hyperviscosity. If we look at the zonal component of the
ST-ITG spectrum (dashed green curve) it falls off faster between the
peak and the hyperviscous dissipation scale, $k_{\nu,E}\text{(ST)}$.
After $k_{\nu,E}\text{(ST)}$, the zonal spectrum falls off similar
to the non-zonal spectrum due to hyperviscous damping.

In contrast, the total energy spectrum of the PV-ITG model (solid
orange curve) and its zonal component (dashed red curve) peak at a
larger value of $k$ than the ST-ITG model. This is in agreement with
the observation that the zonal flows do not merge {[}figure~\ref{fig:vbar_xt_comp_PV}{]}
in PV-ITG and hence have a higher dominant radial wavenumber. Furthermore,
$k_{\nu,E}\text{(PV)}$ is actually lower than the peak of the spectral
production, $k_{f,E}\text{(PV)}$. This peculiar aspect of the spectrum
is related to the hyperviscous damping of the zonal component of the
PV-ITG model (dashed red curve). Since the zonal flows dominate so
completely in this case, the little amount of dissipation that they
endure determines the total dissipation that the system actually experiences,
which happens at a much earlier wavenumber compared to the non-zonal
part resulting in the total spectrum rising at around $k_{f,E}\text{(PV)}$
before falling again.

For $\kappa_{T}=2.0$ {[}figure~\ref{fig:Ek_comp_kapt_PV}(b){]},
both total energy and the zonal energy spectra for both models closely
track one another, and $k_{f,E}$ and $k_{\nu,E}$ are the same for
both models. The scaling exponent of the spectrum fitted from $k_{f,E}$
to $k_{\nu,E}$, is $-3.24$ for ST-ITG and $-3.34$ for PV-ITG. These
are steeper than the classical forward enstrophy cascade exponent
of $-3$, which is in part due to the limited scale separation between
$k_{f,E}$ and $k_{\nu,E}$ (small $\mathrm{Re}_{\nu}\sim k^{2}_{\nu,E}/k^{2}_{f,E}$)~\citep{Borue_1993}.
On the other hand, the kinetic energy spectrum is close to but steeper
than $-5/3$, the classical inverse cascade exponent, in a limited
domain between $k_{f,E}$ and $k_{\nu,E}$ (see appendix~\ref{appsec:spectra}).
Therefore, in this narrow domain we see signs of the conventional
2D dual cascade spectrum even though the kinetic energy is not conserved.
This difference between the kinetic energy and the total energy is
due to the contribution of the electron density, $\tau\widetilde{\phi}$,
to the latter. Finally, for both models the zonal part of the spectrum
(dashed green and dashed red curves) dominates for very small wavenumbers
as we do not have any hypoviscosity on zonal modes. The only difference
between the spectra is that the zonal component of the PV-ITG model
is higher than that of the ST-ITG model. The spectra of internal energy
from pressure, $E_{P}(k),$ and the generalized total energy, $G_{\mathrm{tot}}(k)$,
for the ST-ITG model and the spectrum of potential enstrophy, $W(k)$,
for the PV-ITG model are shown in appendix~\ref{appsec:spectra}.

\subsection{Spectral flux\label{subsec:Spectral-flux}}

Taking the fluid ITG as a turbulence model, we can compute the evolution
of the total energy spectrum (and other conserved quantities) by averaging
the modal energy budget over a shell of width $\Delta k$~\citep{Frisch_1995,Alexakis_2018},
yielding
\begin{equation}
\partial_{t}E_{\mathrm{tot}}(k,t)+\partial_{k}\Pi_{E}(k,t)=f_{E}(k,t)-d^{(\nu)}_{E}(k,t)-d^{(H)}_{E}(k,t)\,\text{,}\label{eq:energy_spectrum_evolution}
\end{equation}

\noindent where the instantaneous total energy flux through wavenumber
$k$ is 
\begin{equation}
\begin{aligned}\Pi_{E}(k,t) & =\Re\left(\sum_{\left|\boldsymbol{q}\right|\le k}\phi^{*}_{\boldsymbol{q}}\{\phi,(\tau\widetilde{\phi}-\nabla^{2}\phi)\}_{\boldsymbol{q}}\right)+\Re\left(\sum_{\left|\boldsymbol{q}\right|\le k}\phi^{*}_{\boldsymbol{q}}[\boldsymbol{\nabla}\cdot\{\boldsymbol{\nabla}\phi,P\}]_{\boldsymbol{q}}\right)\\
 & =\Pi^{(\phi)}_{E}(k,t)+\Pi^{(d)}_{E}(k,t)\,\text{.}
\end{aligned}
\label{eq:spectral_flux_E}
\end{equation}

Here, $\Pi^{(\phi)}_{E}(k,t)$ and $\Pi^{(d)}_{E}(k,t)$ are the instantaneous
energy fluxes due to the $\mathrm{E}\times\mathrm{B}$ and the diamagnetic
nonlinearities respectively. If $\Pi_{E}(k,t)$ is positive (and roughly
constant as a function of $k$), it implies a forward cascade, and
if negative, an inverse cascade. The spectral production, spectral
hyperviscous, and hypoviscous dissipations are
\begin{align}
f_{E}(k,t) & =\frac{1}{\Delta k}\Re\left(\sum_{k<\left|\boldsymbol{q}\right|\le k+\Delta k}\kappa_{B}iq_{y}\phi^{*}_{\boldsymbol{q}}P_{\boldsymbol{q}}\right)\,\text{,}\label{eq:spectral_production_E}\\
d^{(\nu)}_{E}(k,t) & =\frac{1}{\Delta k}\sum_{k<\left|\boldsymbol{q}\right|\le k+\Delta k}\nu q^{6}(\tau\mathds{1}_{q_{y}\neq0}+q^{2})|\phi_{\boldsymbol{q}}|^{2}\,\text{,}\label{eq:spectral_dissipation_nu_E}\\
d^{(H)}_{E}(k,t) & =\frac{1}{\Delta k}\sum_{k<\left|\boldsymbol{q}\right|\le k+\Delta k}Hq^{-4}(\tau\mathds{1}_{q_{y}\neq0}+q^{2})|\widetilde{\phi}_{\boldsymbol{q}}|^{2}\,\text{.}\label{eq:spectral_dissipation_H_E}
\end{align}

\begin{figure*}
\centering{}\includegraphics{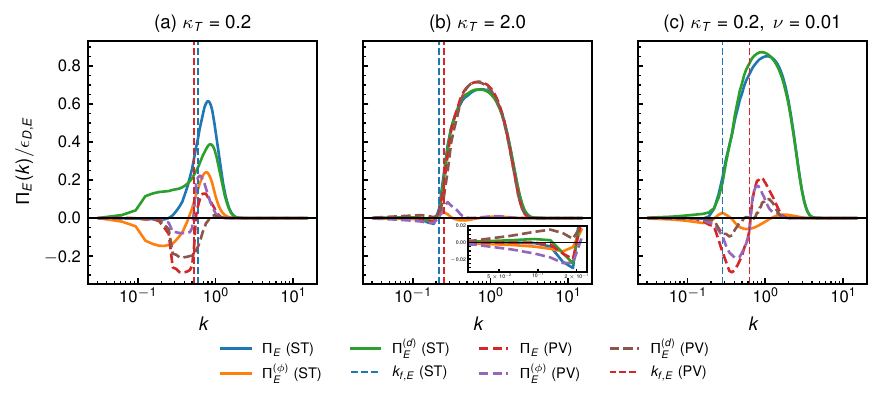}\caption{Spectral flux of the total energy, $\Pi_{E}(k)$, normalized by the
dissipation, $\epsilon_{D,E}$. (a) $\kappa_{T}=0.2$: $\Pi^{(d)}_{E}$
of the ST-ITG model is always positive unlike the PV-ITG model where
it is mostly negative except in a small wavenumber domain around $k=1$
and at $k<0.2$, for ST-ITG, $\Pi^{(d)}_{E}$ cancels out $\Pi^{(\phi)}_{E}$.
In addition, the PV-ITG flux is negative and positive below and above
$k_{f,E}\text{(PV)}$ respectively. (b) $\kappa_{T}=2.0$: $\Pi^{(d)}_{E}$
is dominant and results in a transfer of energy to larger wavenumbers.
(c) $\kappa_{T}=0.2$ and $\nu=0.01$: The ST-ITG flux is dominated
by $\Pi^{(d)}_{E}$ and is positive up to $k\approx4$, after which
it reduces to zero. The PV-ITG fluxes look the same as in (a), with
the difference being that the positive maximum of the diamagnetic
component is higher than that of $\nu=0.1$.\label{fig:Ek_flux_comp_kapt_PV_nu}}
\end{figure*}

The spectral total energy fluxes, averaged over the second half of
the simulation, $\Pi_{E}(k)$, of both models are shown in figure~\ref{fig:Ek_flux_comp_kapt_PV_nu}
for $\kappa_{T}=0.2$ and $\kappa_{T}=2.0$. The wavenumber corresponding
to the peak of $f_{E}(k)$, $k_{f,E}$, is shown as a vertical line
for both models. The condition for steady state of equation~(\ref{eq:energy_spectrum_evolution}),
$\partial_{k}\Pi_{E}(k)=f_{E}(k)-d^{(\nu)}_{E}(k)-d^{(H)}_{E}(k)$,
can be integrated over $k$ to show that the spectral flux has to
balance the production ($\epsilon_{f,E}$) minus the dissipation ($\epsilon_{D,E}=\epsilon^{(\nu)}_{D,E}+\epsilon^{(H)}_{D,E}$).
The fluxes shown in the figure are normalized by the dissipation.

For $\kappa_{T}=0.2$, the fluxes for ST-ITG (solid blue curve) and
the PV-ITG (dashed red curve) are initially negative but start to
differ near $k\approx0.2$, where the ST-ITG flux increases and becomes
positive, while the PV-ITG flux decreases and becomes more negative.
The ST-ITG flux continues to increase, peaks after $k_{f,E}\text{(ST)}=0.595$,
and goes down to zero. The PV-ITG flux reaches its minimum, begins
to rise around $k_{f,E}\text{(PV)}=0.532$, peaks at $k\approx0.7$,
and eventually drops to zero. As with heat transport, the magnitude
of the positive peak of ST-ITG is higher than that of PV-ITG. The
PV-ITG system has pronounced negative and positive fluxes below and
above $k=k_{f,E}$, respectively, consistent with the directions of
energy transfers in 2D turbulence driven at mid-range scales. However,
these fluxes are not sufficiently constant to constitute a dual cascade.

We can also look at the $\mathrm{E}\times\mathrm{B}$ and diamagnetic
components of these fluxes separately. The $\mathrm{E}\times\mathrm{B}$
component of the energy flux for ST-ITG (solid orange curve) is negative
up to $k=0.450$, after which it becomes positive, peaks at $k=0.751$,
and then goes to zero. The diamagnetic component of the flux (solid
green curve), which is positive for all $k$, largely cancels the
$\mathrm{E}\times\mathrm{B}$ component for $k<0.2$, resulting in
an extremely small but negative (i.e., inverse) flux for large scales.
Beyond the forcing scale, both components are positive and add up
to result in a forward transfer of total energy to small scales.

For the PV-ITG model, both flux components are small and negative,
resulting in an essentially small and negative flux for $k<0.2$.
On the other hand, the fluxes of the PV-ITG model (dashed red curve)
are both negative and significant for $k\in(0.07,0.56)$, with the
diamagnetic component of the flux (dashed brown curve) being larger
in magnitude than the $\mathrm{E}\times\mathrm{B}$ component (dashed
purple curve). Hence, they add up to give a clear inverse energy transfer
at these scales. Whereas, for $k>0.56$ the components of the flux
are opposite in sign with a positive and larger $\mathrm{E}\times\mathrm{B}$
flux and a diamagnetic flux that is mostly negative except for a small
domain around $k=1$.

For $\kappa_{T}=2.0$, the fluxes of both models (solid blue and dashed
red curves) are dominated by the diamagnetic component for $k>k_{f,E}$.
This means that at higher temperature gradients, the diamagnetic term
drives a strong forward cascade of total energy, which results in
a turbulent state as a balance between injection and dissipation,
reminiscent of the Kolmogorov cascade picture of turbulence. The $\mathrm{E}\times\mathrm{B}$
components of both fluxes (solid orange and dashed purple curves)
are similar with the only difference being that, for $k>k_{f,E}$,
the $\mathrm{E}\times\mathrm{B}$ component of the ST-ITG flux dips
into the negative flux range while the $\mathrm{E}\times\mathrm{B}$
component of the PV-ITG flux stays positive. Furthermore, as $\kappa_{T}$
increases from $0.2$ to $2.0$, there is a reduction in $k_{f,E}$
from $0.595$ to $0.219$ for ST-ITG and $0.532$ to $0.250$ for
PV-ITG roughly consistent with the mixing length estimates in section~\ref{sec:Linear-Analysis}.

For $\kappa_{T}=2.0$ and $k<0.2$ {[}inset of figure~\ref{fig:Ek_comp_kapt_PV}(b){]},
the diamagnetic component of the ST-ITG spectral flux (solid green
curve) is small and positive, becoming negative before it peaks, whereas
the diamagnetic component of the PV-ITG flux (dashed brown curve)
is small and always positive before peaking. However, the $\mathrm{E}\times\mathrm{B}$
component for both models is negative and is larger in magnitude in
regions where the diamagnetic component is positive resulting in an
extremely small net inverse transfer of energy at large scales.

Note that the only case where the diamagnetic component of the flux
is prominently negative is for the PV-ITG system in the case of $\kappa_{T}=0.2$.
In contrast, the diamagnetic component of the flux in ST-ITG is always
positive for $\kappa_{T}=0.2$ and therefore competes against the
inverse cascade at large scales, making it harder to form zonal flows.
Nevertheless, the system forms zonal flows through the $\mathrm{E}\times\mathrm{B}$
Reynolds stress term as it does in the absence of the diamagnetic
term. In contrast, for PV-ITG, the diamagnetic term increases the
efficiency of the inverse transfer while competing against the forward
transfer by the $\mathrm{E}\times\mathrm{B}$ nonlinearity, making
the formation of zonal flows easier. This explains why the zonal flows
are stronger and more robust in the PV-ITG case. It also explains
why if we decrease the hyperviscosity coefficient for $\kappa_{T}=0.2$,
the ST-ITG system switches to a turbulent state while the PV-ITG system
remains dominated by zonal flows (see section~\ref{sec:Viscous-dissipation}).

The spectral fluxes of the internal energy from pressure, $\Pi_{P}(k)$,
and generalized total energy, $\Pi_{G}(k)$, for the ST-ITG model,
and the spectral flux of potential enstrophy, $\Pi_{W}(k)$, for the
PV-ITG model are shown in appendix~\ref{appsec:spectral-fluxes}.

\begin{figure*}
\centering{}\includegraphics{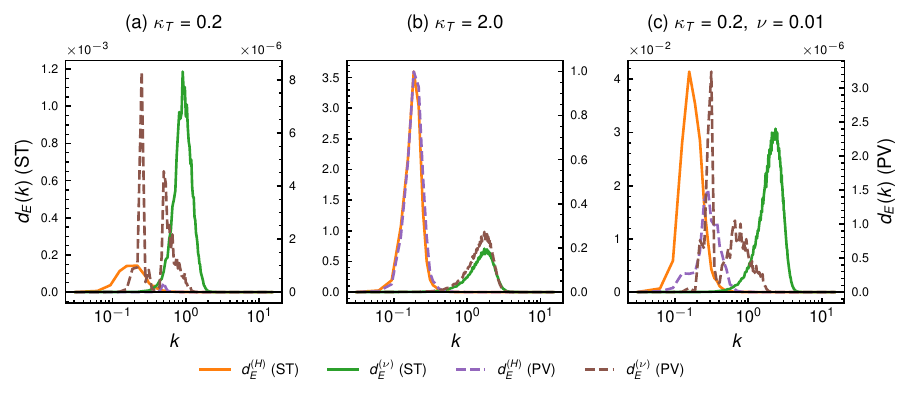}\caption{The hypoviscous and hyperviscous components of the spectral dissipation
of the total energy, $d^{(H)}_{E}(k)$ and $d^{(\nu)}_{E}(k)$, for
the two models. The left y-axes correspond to the spectral dissipations
of ST-ITG and the right y-axes correspond to the spectral dissipations
of PV-ITG. (a) $\kappa_{T}=0.2$: The ST-ITG system's spectral dissipations
are larger in magnitude and $k_{H,E}$ and $k_{\nu,E}$ are separated
by less than a decade. $k_{H,E}$ of PV-ITG is around the secondary
peak of $d^{(\nu)}_{E}\text{(PV)}$. (b) $\kappa_{T}=2.0$: the hypoviscous
component, $d^{(H)}_{E}(k)$, and hyperviscous, $d^{(\nu)}_{E}(k)$,
component of both models look similar and their peaks are separated
by roughly one decade. However, the magnitude of the spectral dissipation
of the ST-ITG model (solid orange and solid green curves) is larger
than that of the PV-ITG model (dashed purple and dashed brown curves).
(c) $\kappa_{T}=0.2$ and $\nu=0.01$: $k_{H,E}$ and $k_{\nu,E}$
for ST-ITG are separated by more than a decade. The PV-ITG spectral
dissipations (dashed) are small in magnitude compared to the ST-ITG
spectral dissipations (solid). Moreover $k_{H,E}\text{(PV)}$ is lower
than the primary peak of $d^{(\nu)}_{E}(k)\text{(PV)}$.\label{fig:Ek_dissipation_comp_kapt_PV_nu}}
\end{figure*}

The spectral dissipations of total energy for $\kappa_{T}=0.2$ and
$\kappa_{T}=2.0$ are shown for both models in figure~\ref{fig:Ek_dissipation_comp_kapt_PV_nu}.
For $\kappa_{T}=0.2$ {[}figure~\ref{fig:Ek_dissipation_comp_kapt_PV_nu}(a){]},
the hyperviscous component of ST-ITG (solid green curve) is dominant
and there is extremely limited scale separation (less than a decade)
between its peak and that of the hypoviscous component. Furthermore,
the maximum of $d^{(\nu)}_{E}(k)$ is roughly two orders of magnitude
larger than that of the PV-ITG counterpart. For PV-ITG, the hyperviscous
component (dashed brown curve) is larger than the hypoviscous one
(dashed purple curve). The hyperviscous component has two peaks corresponding
to the peaks of the zonal and the non-zonal total energy spectra for
the PV-ITG system {[}figure~\ref{fig:Ek_comp_kapt_PV}(a){]}, and
the peak of the hypoviscous component coincides with the secondary
peak of the hyperviscous spectral dissipation. This means that in
this case, the intermediate scale acts both as a small-scale damping
for zonal flows and as a large-scale damping for the fluctuations.

For $\kappa_{T}=2.0$ {[}figure~\ref{fig:Ek_dissipation_comp_kapt_PV_nu}(b){]},
both models display similar spectral dissipations differing only in
magnitude, with the spectral dissipation of the ST-ITG system being
larger than the PV-ITG system. The peak of the hypoviscous spectral
dissipation, $k_{H,E}$, is lower than $k_{\nu,E}$ and the two are
separated by roughly one decade.

\subsection{Instability assumption\label{subsec:Instability-assumption}}

To further study the direction of energy cascade, we use Waleffe's
approach of ``Instability assumption''. This approach relies on
the argument that the direction of the energy transfer can reasonably
be deduced from the stability characteristics of the elementary triad
interactions~\citep{Waleffe_1992,Waleffe_1993}. More precisely,
if there is a clear choice where one wavenumber is unstable, the energy
flows from the unstable wavenumber to the other two. Moreover, this
is also true statistically in a turbulent steady state. To study the
stability characteristics of an elementary triad, consider a case
in which one of the modes in a triad $\boldsymbol{k}+\boldsymbol{p}+\boldsymbol{q}=0$
is much more populated than the others. Assuming $k<p<q$, we take,
without loss of generality, $\left|P_{k0}\right|,\left|\phi_{k0}\right|\gg\left|P_{q0}\right|,\left|\phi_{q0}\right|,\left|P_{p0}\right|,\left|\phi_{p0}\right|$,
to obtain a linear system that can be solved for the growth rate,
$\lambda$. The linear systems for the other modes as pumps can be
obtained by cyclic permutations of $(k,p,q)$. Generally, when a 2D
system is comprised solely of the Reynolds stresses like in hydrodynamic
turbulence~\citep{Fjortoft_1953} or drift-wave turbulence~\citep{Hasegawa_1978,Hasegawa_1979}
we observe the following:
\begin{itemize}
\item When the middle mode, $p$, is dominant, there is an instability resulting
in both $k$ and $q$ modes growing.
\item When either $k$ or $q$ is dominant, the system is stable.
\end{itemize}
Hence, the spectrum cascades by the simultaneous excitation of the
longer and shorter wavelengths which allows us to conclude that there
is a dual cascade of the two conserved quantities: energy and enstrophy.
And a Fjørtoft argument, which uses the fact that the enstrophy, $W_{k}$,
is $k^{2}$ times the energy, $E_{k}$, allows us to conclude that
the energy cascades towards smaller $k$ while the enstrophy cascades
towards larger $k$. 

The triadic conservation of the internal energy from pressure (ST-ITG),
total energy (both models), generalized total energy (ST-ITG), and
potential enstrophy (PV-ITG) implies
\begin{align}
\Delta\left|P_{q}\right|^{2}+\Delta\left|P_{k}\right|^{2}+\Delta\left|P_{p}\right|^{2} & =0\,\text{,}\label{eq:Delta_EP}\\
(\tau+q^{2})\Delta\left|\phi_{q}\right|^{2}+(\tau+k^{2})\Delta\left|\phi_{k}\right|^{2}+(\tau+p^{2})\Delta\left|\phi_{p}\right|^{2} & =0\,\text{,}\label{eq:Delta_E}\\
(\tau+q^{2})\Delta\left|\phi_{q}+P_{q}\right|^{2}+(\tau+k^{2})\Delta\left|\phi_{k}+P_{k}\right|^{2}+(\tau+p^{2})\Delta\left|\phi_{p}+P_{p}\right|^{2} & =0\,\text{,}\label{eq:Delta_G}\\
\Delta\left|-(\tau+q^{2})\phi_{q}+P_{q}/\Gamma\right|^{2}+\Delta\left|-(\tau+k^{2})\phi_{k}+P_{k}/\Gamma\right|^{2}+\Delta\left|-(\tau+p^{2})\phi_{p}+P_{p}/\Gamma\right|^{2} & =0\,\text{,}\label{eq:Delta_W}
\end{align}
where $\Delta f=f_{t}-f_{0}$ is the difference of $f$ at time $t$
and $f$ at time $t=0$. For the ST-ITG system, since, both total
energy and generalized total energy scale similarly in wavenumber,
the spectral transfer is not governed by a traditional $E_{k}$ vs.
$k^{2}E_{k}$ (energy vs. enstrophy) budget~\citep{Fjortoft_1953}
that forces a forward cascade of enstrophy and inverse cascade of
kinetic energy. A change in $|\phi_{k}|^{2}$ cannot exist without
a compensatory change in the cross term, $\Re(\phi_{k}P^{*}_{k})$,
and $k^{2}P^{2}$ to preserve the conservation of generalized total
energy. Effectively, the nonlinearity shuffles energy between the
electric field and pressure fluctuations. The phase difference between
the pressure and the potential, $\delta_{k}$, provides an extra degree
of freedom. By adjusting $\delta_{k}$ across scales, the system can
satisfy the conservation laws even while the energy cascades forward.
This explains why the inclusion of diamagnetic nonlinearity allows
the system to deviate from the typical inverse energy cascade prediction
of 2D flows.

For the PV-ITG system, where it is the total energy and the potential
enstrophy that are conserved, the potential enstrophy equation~(\ref{eq:Delta_W})
results in the higher order term $(\tau+k^{2})^{2}\phi^{2}_{k}$ but
it still has $P^{2}_{k}$ and a cross term, $\Re([\tau+k^{2}]\phi_{k}P^{*})$.
For large wavenumbers, due to the $(\tau+k^{2})^{2}\phi^{2}_{k}$
term, which scales as $k^{4}\phi^{2}_{k}$, one expects the traditional
picture of forward cascade of enstrophy to work. For medium and small
wavenumbers, we also expect the $k^{2}\phi^{2}_{k}$ term and cross-terms
to contribute to the potential enstrophy budget allowing for the potential
enstrophy to be nonlinearly transferred to smaller wavenumbers without
breaking total energy conservation (see appendix~\ref{appsec:spectral-fluxes}).

\begin{figure*}
\begin{centering}
\includegraphics{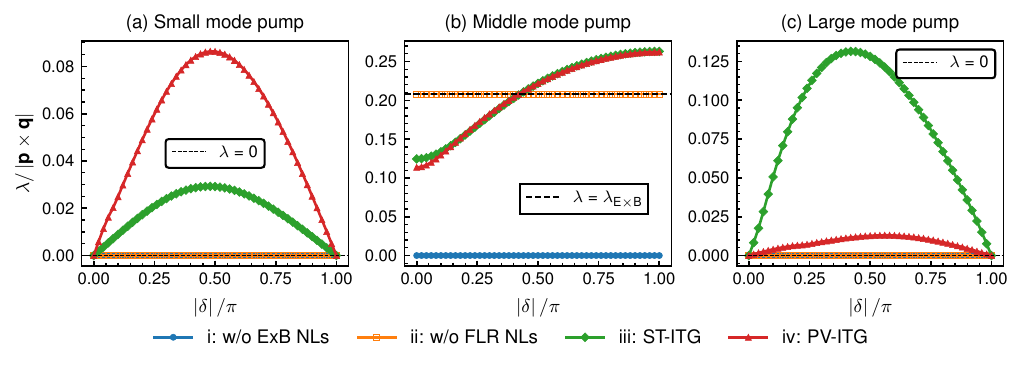}
\par\end{centering}
\caption{Growth rate, $\lambda$, normalized by the area of the triad, $|\boldsymbol{p}\times\mathbf{q}|$,
vs. the absolute value of the pump mode phase difference, $|\delta|$,
for different nonlinearity cases for the triad \textbf{$\boldsymbol{k}=\left(0.3,0.3\right),\boldsymbol{p}=\left(0,0.6\right),\boldsymbol{q}=\left(-0.3,-0.9\right)$}.
As seen by the blue curve, the FLR nonlinearities alone cannot make
the triad interactions unstable. On the other hand, the combined presence
of FLR nonlinearities together with the $\mathrm{E}\times\mathrm{B}$
nonlinearities makes the triad unstable for both small- and large-mode
pumps (green and red). (a) Small-mode pump: The growth rate of the
PV-ITG model is larger than that of the ST-ITG model. (b) Middle-mode
pump: The growth rate of the system without FLR nonlinearities (orange)
is finite and is $|\delta|$ independent. The growth rates of the
ST-ITG and PV-ITG models are always positive and they increase with
$|\delta|$. (c) Large-mode pump: the growth rate of the ST-ITG is
higher than that of the PV-ITG model.\label{fig:Growth-rate-nonzonal}}
\end{figure*}

When we introduce the FLR nonlinearities, drift-wave potential enstrophy
is no longer conserved and it is the generalized total energy (ST-ITG)
and potential enstrophy (PV-ITG) that are conserved resulting in a
change in the behavior of the 2D cascade. We would like to observe
this change in the instability characteristics of a triad and to see
how well it is related to the cascade, as described in the earlier
sections. To do so, we look at the following cases with different
nonlinearities to better understand the role of the FLR nonlinearities
for the triad $\boldsymbol{k}=(0.3,0.3),\boldsymbol{p}=(0,0.6),\boldsymbol{q}=(-0.3,-0.9)$.
The growth rates, $\lambda$, are plotted against the phase difference,
$\delta$, between the pump mode's pressure and potential, $P_{k0}=\phi_{k0}e^{i\delta}$.
\begin{enumerate}[label=(\roman*)]
\item Without $\mathrm{E}\times\mathrm{B}$ nonlinearities (blue curve
with point markers in figure~\ref{fig:Growth-rate-nonzonal}): the
system is stable irrespective of which mode is the pump mode. We conclude
that the FLR nonlinearities alone cannot result in spectral energy
transfer. 
\item Without FLR nonlinearities (orange curve with open-square markers
in figure~\ref{fig:Growth-rate-nonzonal}): the system only has $\mathrm{E}\times\mathrm{B}$
advection nonlinearities and so is unstable when the middle mode is
the pump and stable otherwise, resulting in the standard dual cascade
picture.
\item ST-ITG (green curve with diamond markers in figure~\ref{fig:Growth-rate-nonzonal}):
The system is unstable for all 3 pump modes which is a deviation from
case (ii). For middle-mode pump (center plot) the growth rate is lowest
at $\left|\delta\right|=0$ and monotonically increases and flattens
at $\left|\delta\right|=\pi$. Note that this variation with $\delta$
is about a mean value that is the typical growth rate of the $\mathrm{E}\times\mathrm{B}$
advection system (case ii). For the small-mode and large-mode pumps
{[}(a) and (c) respectively{]}, the growth rate maximizes at a finite
$\left|\delta\right|<\pi/2$ and drops to zero at $\left|\delta\right|=0,\pi$.
\item PV-ITG (red curve with triangle markers in figure~\ref{fig:Growth-rate-nonzonal}):
The system is unstable for all 3 pump modes as in the ST-ITG case
{[}(iii){]}. The growth rate of the PV-ITG curve is higher than that
of ST-ITG (green) for the small mode pump {[}(a){]}, similar but slightly
lower for the middle-mode pump {[}(b){]} and lower than that of ST-ITG
for the large-mode pump {[}(c){]}. This suggests that the PV-ITG system
is likely to result in more nonlinear transfer of energy to large
wavenumbers (forward cascade) than the ST-ITG system. If we scale
up the wavenumbers (say by multiplying each wavenumber by 5), then
the PV-ITG growth rate is much higher than the ST-ITG growth rate
for small-mode and middle-mode pump (still highest among the three
pumps) while for the large-mode pump, the PV-ITG growth rate is now
only slightly bigger than the ST-ITG growth rate. This indicates that,
for larger wavenumbers, the PV-ITG system is likely to transfer more
energy to larger wavenumbers compared to the ST-ITG system. This might
explain the higher $\Pi_{E}(k)/\epsilon_{E}$ observed for the PV-ITG
system for $\kappa_{T}=2.0$ {[}figure~\ref{fig:Ek_flux_comp_kapt_PV_nu}{]}.
\end{enumerate}
The phase dependency of the growth rate, in the presence of the FLR
nonlinearities {[}case (iii) and case (iv){]}, is consistent with
the cross-terms between pressure and potential entering the conservation
budget of the generalized total energy, $G_{\mathrm{tot}}$, and potential
enstrophy, $W$. Triads with the injection mode as the lowest wavenumber
have a larger nonlinear coefficient and hence larger growth rate,
$\lambda$, than triads with the injection mode as the middle or highest
wavenumber. Consequently, the additional instability characteristics
of the triad for small-mode and large-mode pumps indicate an additional
forward energy cascade. This is visible in the 2D simulation in the
form of small-scale vortices (seen in inset of figure~\ref{fig:hypo_comp})
and in the positive contribution of the diamagnetic component of the
spectral flux of energy in figure~\ref{fig:Ek_flux_comp_kapt_PV_nu}.

\section{Role of viscous dissipation\label{sec:Viscous-dissipation}}

In this section we would like to study the role of the hyperviscosity
coefficient on the system. The hyperviscosity is required because
the FLR nonlinearities result in a spectrum that is steeper than what
is generally found for conventional 2D systems and we find that it
is necessary for the saturation of the PV-ITG simulation. It also
has the nicer advantage that it reduces the amplitude of the bursts
in the ST-ITG evolution compared to using Laplacian viscosity. However,
even in situations where hyperviscosity is used, we find that we have
to use extremely high values of the viscosity coefficient. This is
due to the requirement of eliminating the dynamics of medium to large
wavenumbers from the system. However, finally, we end up increasing
the magnitude of the hyperviscosity coefficient even beyond the magnitude
required for saturation, $\nu=0.01$, due to the dependence of the
zonal flow levels of the ST-ITG system on the hyperviscosity coefficient.
Here we intend to clarify this issue.

The zonal kinetic energy fraction and box-averaged heat flux of both
the models for $\kappa_{T}=0.2$ with $\nu=0.01$ are shown in figures~\ref{fig:zonal_KE_frac_Q_vs_t_comp_kapt_PV}(c)
and~(d) respectively. We observe that $\langle\Xi\rangle_{T/2}$
of the ST-ITG curve (cyan) is $0.117$, which is lower than that of
the $\nu=0.1$ curve (blue) in figure~\ref{fig:zonal_KE_frac_Q_vs_t_comp_kapt_PV}(a),
$\langle\Xi\rangle_{T/2}=0.926$. The evolution is also more bursty
compared to the $\nu=0.1$ curve. However, $\langle\Xi\rangle_{T/2}$
of the PV-ITG simulation for $\nu=0.01$ (brown) is $0.995$ which
is roughly the same as that of the $\nu=0.1$ simulation (green) in
figure~\ref{fig:zonal_KE_frac_Q_vs_t_comp_kapt_PV}(b), $\langle\Xi\rangle_{T/2}=0.996$.
The difference is that the zonal kinetic energy fraction of the $\nu=0.01$
simulation takes slightly longer to saturate compared to the $\nu=0.1$
simulation. Moreover, the zonal flows merge between $\gamma t=600$
and $\gamma t=700$ just like the $\nu=0.1$ simulation {[}figure~\ref{fig:vbar_xt_comp_PV}(b){]}.
However, the number of zonal peaks of the $\nu=0.01$ simulation is
one higher than that of the $\nu=0.1$ simulation: $10$ and $9$
before merger whereas $9$ and $8$ after merger respectively. $\langle Q_{\mathrm{box}}\rangle_{T/2}$
of the ST-ITG simulation (cyan) is $5.535$ which is two orders of
magnitude higher than that of the $\nu=0.1$, $\kappa_{T}=0.2$ simulation,
$\langle Q_{\mathrm{box}}\rangle_{T/2}=0.069$. This corresponds well
with the reduction in the zonal kinetic energy fraction suggesting
that the nonlinear threshold has essentially shifted below $\kappa_{T}=0.2$
with the decrease in the hyperviscosity coefficient. On the other
hand, $\langle Q_{\mathrm{box}}\rangle_{T/2}$ of the PV-ITG simulation
(brown) is zero, up to three decimal digits, just like the $\nu=0.1$
simulation (green) suggesting that the PV-ITG nonlinear threshold
has not decreased enough for the system to be in the high-transport
state.

The spectral flux of total energy and spectral dissipation of total
energy of both the models for $\kappa_{T}=0.2$ and $\nu=0.01$ are
shown in figure~\ref{fig:Ek_flux_comp_kapt_PV_nu}(c) and figure~\ref{fig:Ek_dissipation_comp_kapt_PV_nu}(c)
respectively. The ST-ITG flux of $\nu=0.01$ has widened compared
to $\nu=0.1$ {[}figure~\ref{fig:Ek_flux_comp_kapt_PV_nu}(a){]}
and is positive up to $k\approx4$. Moreover, it is now dominated
by the diamagnetic component. The widening is due to the increased
scale separation between the $k_{H,E}$ and $k_{\nu,E}$ as a result
of the reduction in $\nu$. So, there is an extended domain for forward
cascade, which might cause more energy to be transferred out of low-wavenumber
modes, leaving less energy available to zonal flows.

The PV-ITG flux of $\nu=0.01$ is qualitatively the same as that of
$\nu=0.1${[}figure~\ref{fig:Ek_flux_comp_kapt_PV_nu}(a){]} with
the only difference being that the diamagnetic component, after $k=k_{f,E}\text{(PV)}$,
rises to a positive value larger than that of the $\nu=0.1$ case.
The wavenumber at which $d^{(H)}_{E}(k)$ maximizes has reduced from
$k=0.5$ (for $\nu=0.1$) to $k=0.280$ (for $\nu=0.01$) {[}figures~\ref{fig:Ek_dissipation_comp_kapt_PV_nu}(a)
and~(c){]}. Furthermore, the $d^{(\nu)}_{E}(k)$ still has two peaks.
The primary peak's position has increased from $k=0.25$ to $k=0.313$
while the secondary peak's position has increased from $k=0.5$ (for
$\nu=0.1$) to $k=0.655$ (for $\nu=0.01$). Therefore, for $\nu=0.01$,
the $k_{H,E}$ is slightly to the left of $k_{\nu,E}$, whereas, for
$\nu=0.1$, $k_{H,E}$ coincides with the secondary peak of $d^{(\nu)}_{E}(k)$.
On the other hand, the $\kappa_{T}=2.0$ behavior remains largely
unchanged and hence is not shown.

We observe that the zonal flow level of the ST-ITG model and scale
separation of the spectral dissipations of the PV-ITG model vary depending
on the coefficient of viscosity, which controls how much the medium
wavenumbers participate in the dynamics of the system. Ideally, the
medium and high $k$ behavior of the FLR nonlinearities would be such
that we don't have to rely so heavily on the dissipation term to capture
the zonal flow physics. We discuss such modifications to the FLR nonlinearities
in section~\ref{sec:Conclusion}.

\section{Conclusion\label{sec:Conclusion}}

We considered simple two dimensional, fluid ITG models consisting
of what can be called a drift-wave potential vorticity advection equation
coupled to a pressure or temperature equation. We found that these
simple systems, without the diamagnetic nonlinearity, always end up
in a zonal flow dominated state. On the other hand, if we keep the
diamagnetic nonlinearity, we obtain a system (ST-ITG) that can describe
the transition from a zonal flow dominated state, to a turbulent state.
However, being a 2D turbulence model, such a system requires large
scale dissipation, which we introduce in the form of hypoviscosity,
in order to saturate. Also because the form of the diamagnetic nonlinearity
that one uses in such fluid models is equivalent to a small wavenumber
Taylor expansion of a gyrofluid model with accurate FLR physics, the
behavior of this nonlinear term becomes pathological at small scales.
Therefore one must also use stronger small scale dissipation, which
we choose to implement in the form of hyperviscosity.

However, the system discussed above does not conserve potential vorticity.
In fact, it may initially appear that the ITG fluid system would not
conserve potential vorticity, because of the curvature and parallel
compression terms. This is true if one uses drift-wave potential vorticity,
which is basically the ion guiding center density. However using a
more general definition of potential vorticity, which follows from
Ertel's theorem~\citep{Gurcan_2015}, one can show that, by keeping
similarly higher order terms in the pressure equation, it is possible
to write down a potential vorticity conserving fluid ITG model (PV-ITG).

Comparing the PV-ITG to the ST-ITG we find that we need hypoviscosity
and hyperviscosity in both models in order to saturate the turbulent
state, with rather high hyperviscosity coefficients so that high wavenumbers,
where the inaccuracies of the FLR expansion of the nonlinearities
start to cause problems, are sufficiently damped. We find also that
the zonal flows are more stationary and their levels are more resilient
to the value of the hyperviscosity coefficient in PV-ITG compared
to ST-ITG, where they are somewhat oscillatory and their dominance
depends on the hyperviscosity coefficient. It was observed, for instance,
that when we decreased the hyperviscosity coefficient tenfold, the
nonlinear threshold of PV-ITG did not decrease below $\kappa_{T}=0.2$
while that of the ST-ITG system did. 

However, these differences cannot be attributed exclusively to potential
vorticity conservation, since the additional terms required to restore
this conservation law also modify the linear and nonlinear dynamics
of the system. A systematic term-by-term investigation of the additional
terms introduced in the PV-ITG pressure equation, together with more
extensive $\kappa_{T}$ scans, would be useful for disentangling these
effects and determining their respective contributions to zonal flow
formation and saturation. We leave these investigations for future
work.

A detailed linear study showed that retaining the FLR terms allows
the growth rate to saturate for high $k$, and makes it such that
for $k_{y}<k_{y,\mathrm{max}}$, it is a finite $k_{x}\neq0$ that
maximizes the growth rate for a given $k_{y}$ and not $k_{x}=0$
as one generally assumes. However, this is only relevant for $\kappa_{T}=0.2$
and the effect for $\kappa_{T}=2.0$ is very small.

We have also shown that by looking at the mixing-length estimate for
turbulent diffusivity, $D_{\mathrm{turb}}=(\gamma/k^{2})_{\mathrm{max}}$,
and requiring that this diffusivity remains finite as $k_{y}\rightarrow0$,
one can deduce the need to introduce a hypoviscosity term, before
running any nonlinear simulations. This necessity is also supported
by a nonlinear simulation, with and without hypoviscosity, showing
that the system manages to saturate to a bursty state for $\kappa_{T}=2.0$
with hypoviscosity at large scales and Laplacian viscosity at small
scales. The unrealistically huge bursts, which show signs of being
numerical artifacts, can further be eliminated by using hyperviscosity,
also consistent with the state of the art in gyrofluid models~\citep{Grander_2024}.
Therefore, we conclude that in order to study fluid ITG with zonal
flow destabilization in 2D, one has to choose the proper form of small
and large scale dissipations.

We have also performed comparisons of the wavenumber spectra, even
though the number of decades that one can cover had to be small because
of the FLR terms. In the zonal dominated PV-ITG, the non-zonal part
of the total energy spectrum peaks after the zonal part resulting
in two visible peaks in the spectrum. In the turbulent case, the spectra
of both models look similar and have power law indices slightly steeper
than $k^{-3}$, the classical forward enstrophy cascade spectrum.
However, the spectrum of kinetic energy on a tighter domain is slightly
steeper than but closer to $k^{-5/3}$, the classical inverse cascade
spectrum. The spectra show signs of the conventional dual cascade
spectrum for the kinetic energy even though it is not a conserved
quantity.

Considering the spectral fluxes of the total energy, it is observed
that the diamagnetic nonlinearity results in a modest contribution
for $\kappa_{T}=0.2$ while it dominates for $\kappa_{T}=2.0$. For
$\kappa_{T}=0.2$, the diamagnetic component of the ST-ITG simulation
is always positive while that of the PV-ITG simulation is mostly negative.
However, for $\kappa_{T}=2.0$, the diamagnetic components of the
fluxes for both models are positive and dominate the total flux with
the PV-ITG flux being slightly higher compared to the ST-ITG flux.

The spectral fluxes are also supported by an analysis of the nonlinearities
using Waleffe's ``Instability assumption'' approach (subsection~\ref{subsec:Instability-assumption}).
It is found that the triads are unstable in the presence of FLR nonlinearities
for all pump modes and the growth rates have a dependence on the phase
difference between the pressure and the potential. This means that
a certain triad with a certain cross-phase would grow faster or slower
depending on which mode acts as the pump. We find in particular that
the small- and large-mode pumps are unstable. Although the growth
rates increase with the magnitude of the wavenumbers for all pump
modes, this increase in growth rate is higher for the small- and middle-mode
pumps than for the large-mode pump. Therefore, we expect a net transfer
of energy to larger wavenumbers, as observed for $\kappa_{T}=2.0$.
We conclude that, while it gives a general idea about the direction
of the cascade implied by certain forms of triads, lacking universal
results, it is difficult to extract any information about the different
global behaviors of the system in different temperature gradient regimes,
using this approach.

Note finally that, having established the form of potential vorticity
to be conserved for ITG, we can introduce an additional equation for
the parallel velocity, and write a 3D potential vorticity conserving
ITG model (PV-ITG) as, 
\begin{align}
\partial_{t}P+\{\phi,P\}+\Gamma\boldsymbol{\nabla}\cdot\{\boldsymbol{\nabla}\phi,P\}+(\kappa_{n}+\kappa_{T})\partial_{y}\phi+\Gamma(\kappa_{n}+\kappa_{T})\partial_{y}\nabla^{2}\phi-\Gamma\kappa_{B}\partial_{y}P & =-\Gamma\nabla_{\parallel}v_{\parallel}+D_{P}\,\text{,}\label{eq:itg3d_pv-1}\\
\partial_{t}v_{\parallel}+\{\phi,v_{\parallel}\} & =-\nabla_{\parallel}(\phi+P)+D_{v}\,\text{,}\label{eq:itg3d_pv-2}\\
\partial_{t}(\tau\widetilde{\phi}-\nabla^{2}\phi)+\{\phi,(\tau\widetilde{\phi}-\nabla^{2}\phi)\}+\boldsymbol{\nabla}\cdot\{\boldsymbol{\nabla}\phi,P\}+\kappa_{n}\partial_{y}\phi+(\kappa_{n}+\kappa_{T})\partial_{y}\nabla^{2}\phi-\kappa_{B}\partial_{y}P & =-\nabla_{\parallel}v_{\parallel}+D_{\phi}\,\text{,}\label{eq:itg3d_pv-3}
\end{align}
assuming $k_{\parallel}/\kappa_{n}\sim\mathcal{O}(\epsilon)$ and
no background parallel velocity gradient, that conserves the same
generalized potential vorticity, $\text{PV}=\nabla^{2}\phi-\tau\widetilde{\phi}+P/\Gamma+(\kappa_{n}-[\kappa_{n}+\kappa_{T}]/\Gamma)x$
as the 2D PV-ITG model {[}equations~(\ref{eq:itg2d_pv-1}) and~(\ref{eq:itg2d_pv-2}){]}.
This is possible since the $\Gamma\nabla_{\parallel}v_{\parallel}$
in the pressure equation cancels with the $\nabla_{\parallel}v_{\parallel}$
in the $\phi$ equation. In the slab limit, $\kappa_{B}=0$, this
model is a potential vorticity conserving slab-ITG system similar
to reference~\citealp{Wang_2012}. That work assumes $(k_{\parallel}/\kappa_{n})^{2}\sim\mathcal{O}(\epsilon)$
(rather than the standard $k_{\parallel}/\kappa_{n}\sim\mathcal{O}(\epsilon)$)
to retain parallel compression in the potential equation but considers
the cold-ion limit to ignore pressure fluctuations. This results in
a system that does not conserve drift-wave potential vorticity defined
as $\nabla^{2}\phi-\phi+\kappa_{n}x$ in the dissipationless limit
and introduces an artificial coupling between potential vorticity
and parallel compression in the potential enstrophy evolution equation.
Therefore, if one desires to retain parallel compression while also
conserving potential vorticity, then it is necessary to include a
pressure equation together with a $-\nabla_{\parallel}P$ term in
the RHS of the parallel velocity equation instead of assuming cold
ions. In such a system parallel compression does not provide an additional
driving term for zonal flow generation.

Regarding the damping of small to medium wavenumbers: Note that the
Hasegawa-Wakatani system, in contrast to our model, behaves nicely
at large wavenumbers, since the vorticity equation basically becomes
the Navier-Stokes equation as $k\rightarrow\infty$. This makes Hasegawa-Wakatani
useful as a 2D model because even though its high $k$ behavior is
not physically accurate, it is perfectly well behaved. The issue with
ITG models is partly due to retaining only one additional term in
the Taylor expansion of a gyrokinetic or a properly designed gyrofluid
model~\citep{Brizard_1992}. If we retain two additional nonlinear
terms, the behavior of the system at small scales might improve. However,
much like a higher-order Taylor expansion of a complicated function
evaluated far from its expansion point, the error at small scales
may even increase as a result. Instead, using a gyrofluid model with
a proper FLR representation directly may allow us to capture $k\gg1$
behavior of ITG more accurately. However, including pressure nonlinearities
in such models is not trivial and involves additional assumptions,
whose accuracy needs to be tested in the current context, which we
leave for future studies. 

Furthermore, although the generalized PV considered here is conserved
by the PV-ITG equations in the dissipationless limit, its relationship
to the gyrofluid invariants derived from full-$F$ formulations, such
as those of reference~\citealp{Madsen_2015}, remains to be established.
Deriving a PV-conserving gyrofluid ITG system directly from the gyrocenter-density
and pressure equations, while consistently accounting for magnetic-field
inhomogeneity~\citep{Hahm_2024}, and establishing its relationship
to the full-$F$ gyrofluid invariants through a consistent reduction
are left for future work.

Note also that the inevitable requirement of hypoviscosity may in
fact be stemming from the use of local gradient-driven simulations
with periodic boundary conditions instead of a flux-driven system
with boundary conditions. This is because, in a flux driven system,
whenever there is a jump in the flux, the profile relaxes, mitigating
the increase and thereby allowing the simulation to saturate eventually.
Therefore, a flux-driven code with fixed boundary conditions should
elucidate the behavior of the FLR nonlinearities better and also improve
the system from a modeling perspective. These extensions, too, are
left for future studies.
\begin{acknowledgments}
The authors warmly thank L. Manfredini, P. L. Guillon and P. Morel
for useful discussions. This work has benefited from a grant managed
by the Agence Nationale de la Recherche (ANR), as part of the program
‘Investissements d’Avenir’ under the reference (ANR-18-EURE-0014)
and has been carried out within the framework of the EUROfusion Consortium,
funded by the European Union via the Euratom Research and Training
Programme (Grant Agreement No 101052200 --- EUROfusion) and within
the framework of the French Research Federation for Fusion Studies.
\end{acknowledgments}

\bibliographystyle{unsrt}
\bibliography{references}

\appendix

\section{Conservation laws of the ST-ITG model\label{appsec:3rd-Conservation-law}}

To derive the conservation equation for the generalized total energy,
$G_{\mathrm{tot}}$, we first compute the evolution of the cross-term
$\langle P(\tau\widetilde{\phi}-\nabla^{2}\phi)\rangle$ by summing
the $\phi$ equation multiplied by $P$ and the $P$ equation multiplied
by $(\tau\widetilde{\phi}-\nabla^{2}\phi)$, taking the average, and
using $\langle P\{\phi,(\tau\widetilde{\phi}-\nabla^{2}\phi)\}\rangle+\langle(\tau\widetilde{\phi}-\nabla^{2}\phi)\{\phi,P\}\rangle=0$,
which gives
\begin{equation}
\begin{aligned}\partial_{t}\langle P(\tau\widetilde{\phi}-\nabla^{2}\phi)\rangle & +\langle P\boldsymbol{\nabla}\cdot\{\boldsymbol{\nabla}\phi,P\}\rangle+\kappa_{n}\langle P\partial_{y}\phi\rangle+(\kappa_{n}+\kappa_{T})\langle P\partial_{y}\nabla^{2}\phi\rangle\\
 & =-\nu\langle2\tau\boldsymbol{\nabla}^{3}\widetilde{\phi}\cdot\boldsymbol{\nabla}^{3}P+2\nabla^{4}P\nabla^{4}\phi\rangle-H\langle2\tau\nabla^{-2}\widetilde{\phi}\nabla^{-2}\widetilde{P}+2\boldsymbol{\nabla}^{-1}\widetilde{\phi}\cdot\boldsymbol{\nabla}^{-1}\widetilde{P}\rangle\,\text{,}
\end{aligned}
\end{equation}

\noindent where the dissipation terms are expressed assuming $D=\nu$
for simplicity but they do not affect the form of the conservation
laws themselves. To cancel the second nonlinear term, which has two
$P$s, two $\nabla$s and a Poisson bracket, we need another nonlinear
term of similar form to cancel it out. Therefore, we take the Laplacian
of the pressure equation, multiply by $P$, take the average, and
use $\langle P\nabla^{2}\{\phi,P\}\rangle=\langle P\boldsymbol{\nabla}\cdot\{\boldsymbol{\nabla}\phi,P\}\rangle$
to obtain
\begin{equation}
-\frac{1}{2}\partial_{t}\langle(\boldsymbol{\nabla}P)^{2}\rangle+\langle P\boldsymbol{\nabla}\cdot\{\boldsymbol{\nabla}\phi,P\}\rangle+(\kappa_{n}+\kappa_{T})\langle P\partial_{y}\nabla^{2}\phi\rangle=\nu\langle(\nabla^{4}P)^{2}\rangle+H\langle(\nabla^{-1}\widetilde{P})^{2}\rangle\,\text{.}
\end{equation}

Subtracting this equation from the previous equation for $\partial_{t}\langle P(\tau\widetilde{\phi}-\nabla^{2}\phi)\rangle$
cancels out the remaining nonlinearity resulting in
\begin{equation}
\begin{aligned}\partial_{t}\langle P(\tau\widetilde{\phi}-\nabla^{2}\phi)+\frac{1}{2}(\boldsymbol{\nabla}P)^{2}\rangle & =-\kappa_{n}\langle P\partial_{y}\phi\rangle-\nu\langle2\tau\boldsymbol{\nabla}^{3}\widetilde{\phi}\cdot\boldsymbol{\nabla}^{3}P+2\nabla^{4}P\nabla^{4}\phi+(\nabla^{4}P)^{2}\rangle\\
 & -H\langle2\tau\nabla^{-2}\widetilde{\phi}\nabla^{-2}\widetilde{P}+2\boldsymbol{\nabla}^{-1}\widetilde{\phi}\cdot\boldsymbol{\nabla}^{-1}\widetilde{P}+(\nabla^{-1}\widetilde{P})^{2}\rangle\,\text{.}
\end{aligned}
\end{equation}

Hence, $\langle P(\tau\widetilde{\phi}-\nabla^{2}\phi)+(\boldsymbol{\nabla}P)^{2}/2\rangle$
is conserved in the absence of background gradients and dissipation.
By adding the equations for $\tau\partial_{t}\langle P^{2}\rangle/2$
and $\partial_{t}\langle\widetilde{\phi}^{2}+(\boldsymbol{\nabla}\phi)^{2}\rangle/2$
to the equation above, we obtain the evolution equation for the generalized
total energy,
\begin{equation}
\begin{aligned}\frac{1}{2}\partial_{t}\langle\tau(\widetilde{\phi}+P)^{2}+(\boldsymbol{\nabla}\phi+\boldsymbol{\nabla}P)^{2}\rangle & =-([1+\tau]\kappa_{n}+\tau\kappa_{T}+\kappa_{B})\langle P\partial_{y}\phi\rangle-\Lambda_{G}\,\text{,}\end{aligned}
\label{eq:consv_st-3}
\end{equation}

\noindent where $\Lambda_{G}=-\nu\langle\tau(\boldsymbol{\nabla}^{3}\widetilde{\phi}+\boldsymbol{\nabla}^{3}P)^{2}+(\nabla^{4}\phi+\nabla^{4}P)^{2}\rangle-H\langle\tau(\nabla^{-2}\widetilde{\phi}+\nabla^{-2}\widetilde{P})^{2}+(\boldsymbol{\nabla}^{-1}\widetilde{\phi}+\boldsymbol{\nabla}^{-1}\widetilde{P})^{2}\rangle$.
Note that $G_{\mathrm{tot}}$ is related to $E_{P}$ and $E_{\mathrm{tot}}$
by $G_{\mathrm{tot}}=\tau E_{P}+E_{\mathrm{tot}}+\langle P(\tau\widetilde{\phi}-\nabla^{2}\phi)+(\boldsymbol{\nabla}P)^{2}/2\rangle$.

The other conserved quantities of the ST-ITG model, $E_{P}$ and $E_{\mathrm{tot}}$,
are trivially obtained as
\begin{align}
\frac{1}{2}\partial_{t}\langle P^{2}\rangle & =-(\kappa_{n}+\kappa_{T})\langle P\partial_{y}\phi\rangle-\Lambda_{E_{P}}\,\text{,}\label{eq:consv_st-1}\\
\frac{1}{2}\partial_{t}\langle\tau\widetilde{\phi}^{2}+(\boldsymbol{\nabla}\phi)^{2}\rangle & =-\kappa_{B}\langle P\partial_{y}\phi\rangle-\Lambda_{E}\,\text{,}\label{eq:consv_st-2}
\end{align}

\noindent where $\Lambda_{E_{P}}=\nu\langle(\boldsymbol{\nabla}^{3}P)^{2}\rangle+H\langle(\nabla^{-2}\widetilde{P})^{2}\rangle$
and $\Lambda_{E}=\nu\langle\tau(\boldsymbol{\nabla}^{3}\widetilde{\phi})^{2}+(\nabla^{4}\phi)^{2}\rangle+H\langle\tau(\nabla^{-2}\widetilde{\phi})^{2}+(\boldsymbol{\nabla}^{-1}\widetilde{\phi})^{2}\rangle$.
Furthermore, the time evolution of the drift-wave potential enstrophy,
$\langle(\tau\widetilde{\phi}-\nabla^{2}\phi)^{2}\rangle/2$, is given
as
\begin{equation}
\begin{aligned}\frac{1}{2}\partial_{t}\langle(\tau\widetilde{\phi}-\nabla^{2}\phi)^{2}\rangle & =\kappa_{B}\langle(\tau\widetilde{\phi}-\nabla^{2}\phi)\partial_{y}P\rangle-\underbrace{\langle(\tau\widetilde{\phi}-\nabla^{2}\phi)\boldsymbol{\nabla}\cdot\{\boldsymbol{\nabla}\phi,P\}\rangle}_{\mathrm{NL-transfer}}\\
 & -\nu\langle[\boldsymbol{\nabla}^{3}(\tau\widetilde{\phi}-\nabla^{2}\phi)]^{2}\rangle-H\langle[\nabla^{-2}(\tau\widetilde{\phi}-\nabla^{2}\widetilde{\phi})]^{2}\rangle\,\text{,}
\end{aligned}
\end{equation}

\noindent and the time evolution of the potential enstrophy, $W=\langle(\nabla^{2}\phi-\tau\widetilde{\phi}+P/\Gamma)^{2}\rangle/2$,
for ST-ITG is given as
\begin{equation}
\begin{aligned}\frac{1}{2}\partial_{t}\langle(\nabla^{2}\phi-\tau\widetilde{\phi}+P/\Gamma)^{2}\rangle & =[\kappa_{n}-(\kappa_{n}+\kappa_{T})/\Gamma]/\Gamma\langle P\partial_{y}\phi\rangle+(\kappa_{n}+\kappa_{T})/\Gamma\langle P\partial_{y}\nabla^{2}\phi\rangle-\kappa_{B}\langle(\nabla^{2}\phi-\tau\widetilde{\phi})\partial_{y}P\rangle\\
 & +\underbrace{\langle(\nabla^{2}\phi-\tau\widetilde{\phi}+P/\Gamma)\boldsymbol{\nabla}\cdot\{\boldsymbol{\nabla}\phi,P\}\rangle}_{\mathrm{NL-transfer}}-\nu\langle(\boldsymbol{\nabla}^{3}q)^{2}\rangle-H\langle(\nabla^{-2}\widetilde{q})^{2}\rangle\,\text{,}
\end{aligned}
\label{eq:consv_pv-2}
\end{equation}

\noindent where $q=(\nabla^{2}\phi-\tau\widetilde{\phi}+P/\Gamma)$
is the perturbed part of the potential vorticity. The diamagnetic
nonlinearity results in a nonlinear transfer term that doesn't allow
for $\langle(\tau\widetilde{\phi}-\nabla^{2}\phi)^{2}\rangle/2$ or
$\langle(\nabla^{2}\phi-\tau\widetilde{\phi}+P/\Gamma)^{2}\rangle/2$
to be conserved in the limit of vanishing gradients and dissipation
for ST-ITG.

\section{Spectra of $E_{\mathrm{kin}}$, $E_{P}$, $G_{\mathrm{tot}}$, $W_{\mathrm{DW}}$
and $W$\label{appsec:spectra} }

\subsection{Spectrum of $E_{\mathrm{kin}}$}

The instantaneous kinetic energy spectrum can be written as
\begin{equation}
E_{\mathrm{kin}}(k,t)=\frac{1}{2\Delta k}\sum_{k-\Delta k/2<\left|\boldsymbol{q}\right|\le k+\Delta k/2}q^{2}|\phi_{\boldsymbol{q}}|^{2}\,\text{.}\label{eq:energy_spectrum-1}
\end{equation}

\begin{figure*}
\centering{}\includegraphics{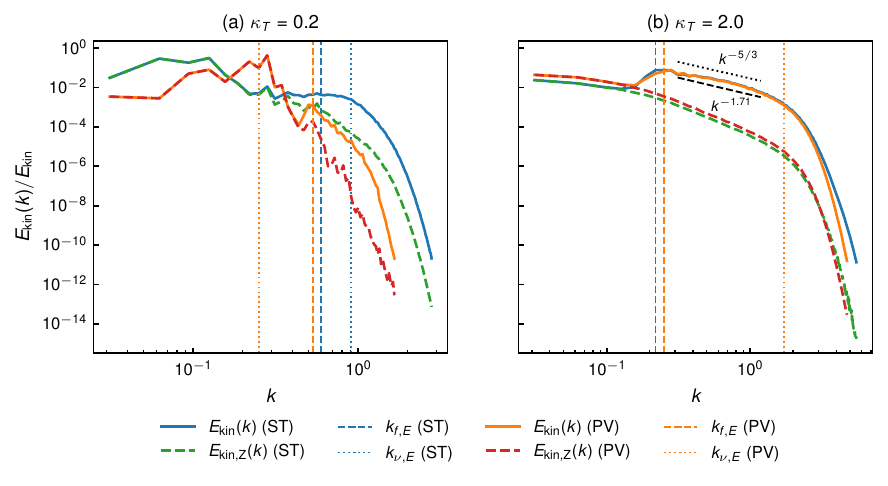}\caption{Normalized kinetic energy spectrum, $E_{\mathrm{kin}}(k)/E_{\mathrm{kin}}$,
vs. the perpendicular wavenumber, $k$, for ST-ITG {[}solid blue and
dashed green curves{]} and PV-ITG {[}solid orange and dashed red curves{]}.
The dashed curves are the zonal components of the spectrum normalized
by the kinetic energy. (a) For $\kappa_{T}=0.2$, the spectra are
identical to the total energy spectra. (b) For $\kappa_{T}=2.0$,
the spectra are generally flatter than the total energy spectra and
are closer to $k^{-5/3}$ in a limited domain enclosed by $(k_{f,E},k_{\nu,E})$,
where $k_{f,E}$ and $k_{\nu,E}$ are forcing and hyperviscous scales
of the total energy.\label{fig:KEk_comp_kapt_PV}}
\end{figure*}

Note that this is the total energy without the contribution from the
electron density, $\tau\widetilde{\phi}$. The spectra, that are time
averaged over the second half of the simulation, are shown in figure~\ref{fig:KEk_comp_kapt_PV}.

The $\kappa_{T}=0.2$ spectra look similar to the total energy spectra
{[}figure~\ref{fig:Ek_comp_kapt_PV}{]} implying that the kinetic
energy is the dominant contribution to the total energy in the zonal-flow-dominated
regime. In contrast, the spectra of $\kappa_{T}=2.0$ are flatter
than the total energy spectra because of the missing contribution
of $\tau\widetilde{\phi}$ as shown in figure~\ref{fig:KEk_comp_kapt_PV}.
Furthermore, the slope of the spectrum for a limited $k$ domain between
$k_{f,E}$ and $k_{\nu,E}$ is close to but greater than $-5/3$ for
$\kappa_{T}=2.0$.

\subsection{Spectrum of $E_{P}$ for ST-ITG}

The instantaneous spectrum of the internal energy from pressure is
defined as
\begin{equation}
E_{P}(k,t)=\frac{1}{2\Delta k}\sum_{k-\Delta k/2<\left|\boldsymbol{q}\right|\le k+\Delta k/2}\left|P_{\boldsymbol{q}}\right|^{2}\,\text{.}
\end{equation}

\begin{figure*}
\centering{}\includegraphics{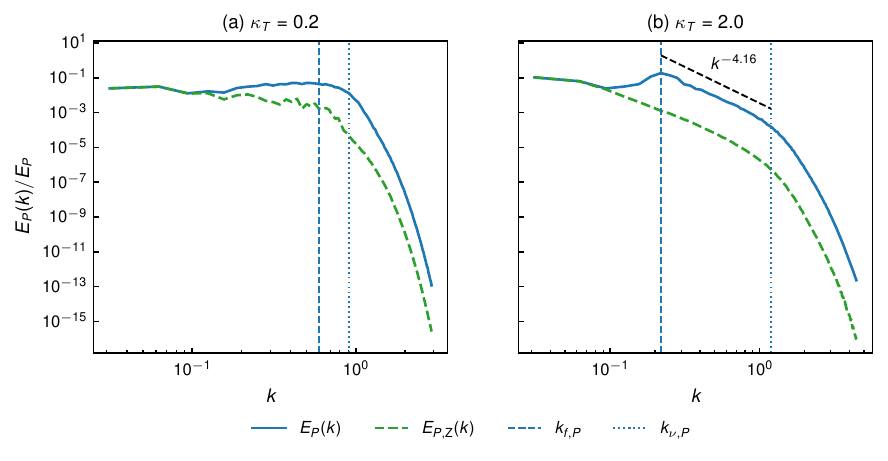}\caption{Normalized spectrum of the internal energy from pressure, $E_{P}(k)/E_{P}$,
vs. the perpendicular wavenumber, $k$, for the ST-ITG model. (a)
For $\kappa_{T}=0.2$, the spectrum is close to being flat for $k<k_{f,P}$.
(b) For $\kappa_{T}=2.0$, the spectrum scales as $E_{P}(k)\sim k^{-4.16}$
in the domain $(k_{f,P},k_{\nu,P})$ and is negatively sloped for
$k<0.1$.\label{fig:EPk_comp_kapt}}
\end{figure*}

The spectra, time averaged over the second half of the simulation,
are shown in figure~\ref{fig:EPk_comp_kapt}. The $\kappa_{T}=0.2$
spectrum is essentially flat for $k<1$ unlike the $\kappa_{T}=2.0$
spectrum which has an exponent of $-4.16$ between $k_{f,P}$ and
$k_{\nu,P}$. 

\subsection{Spectrum of $G_{\mathrm{tot}}$ for ST-ITG}

The instantaneous spectrum of generalized total energy is defined
as
\begin{equation}
G_{\mathrm{tot}}(k,t)=\frac{1}{2\Delta k}\sum_{k-\Delta k/2<\left|\boldsymbol{q}\right|\le k+\Delta k/2}(\tau\mathds{1}_{q_{y}\neq0}+q^{2})|\phi_{\boldsymbol{q}}|^{2}+(\tau+q^{2})\left|P_{\boldsymbol{q}}\right|^{2}+2(\tau\mathds{1}_{q_{y}\neq0}+q^{2})\Re(\phi^{*}_{\boldsymbol{q}}P_{\boldsymbol{q}})\,\text{.}
\end{equation}

\begin{figure*}
\centering{}\includegraphics{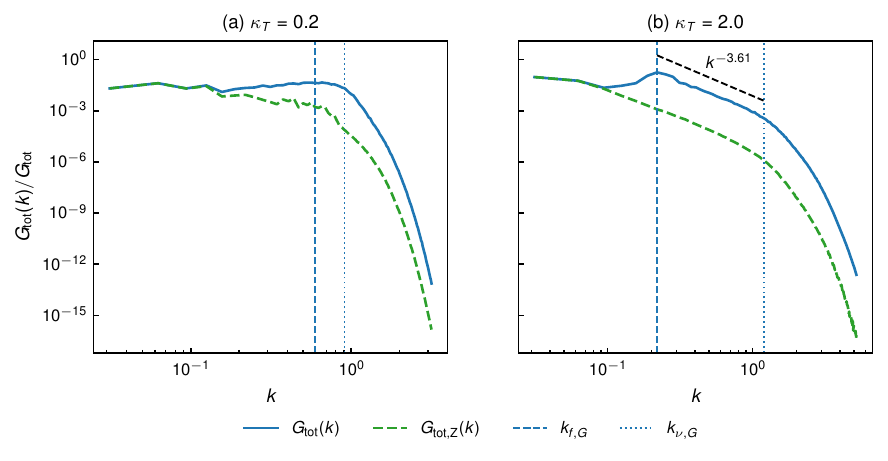}\caption{Normalized generalized total energy spectrum, $G_{\mathrm{tot}}(k)/G_{\mathrm{tot}}$,
vs. the perpendicular wavenumber, $k$, for the ST-ITG model. (a)
For $\kappa_{T}=0.2$, the spectrum is almost flat for $k<k_{f,G}$.
(b) For $\kappa_{T}=2.0$, the spectrum is of the form $G_{\mathrm{tot}}(k)\sim k^{-3.61}$
in the domain $(k_{f,G},k_{\nu,G})$.\label{fig:Gk_comp_kapt}}
\end{figure*}

Note that the first term is $E_{\mathrm{tot}}(k,t)$ and the first
part of the second term is $\tau E_{P}(k,t)$. Therefore, it is the
$q^{2}\left|P_{\boldsymbol{q}}\right|^{2}$ term and the cross-term
that are new. The spectra, time averaged over the second half of the
simulation, are shown in figure~\ref{fig:Gk_comp_kapt}. The $\kappa_{T}=0.2$
spectrum is flat for $k<1$, just like $E_{P}(k)$, unlike the $\kappa_{T}=2.0$
spectrum which has an exponent of $-3.61$ between $k_{f,G}$ and
$k_{\nu,G}$.

\subsection{Spectrum of $W_{\mathrm{DW}}$ for PV-ITG}

The instantaneous drift-wave potential enstrophy spectrum is given
as
\begin{equation}
W_{\mathrm{DW}}(k,t)=\frac{1}{2\Delta k}\sum_{k-\Delta k/2<\left|\boldsymbol{q}\right|\le k+\Delta k/2}(\tau\mathds{1}_{q_{y}\neq0}+q^{2})^{2}|\phi_{\boldsymbol{q}}|^{2}\,\text{.}\label{eq:energy_spectrum-1-1}
\end{equation}

\begin{figure*}
\centering{}\includegraphics{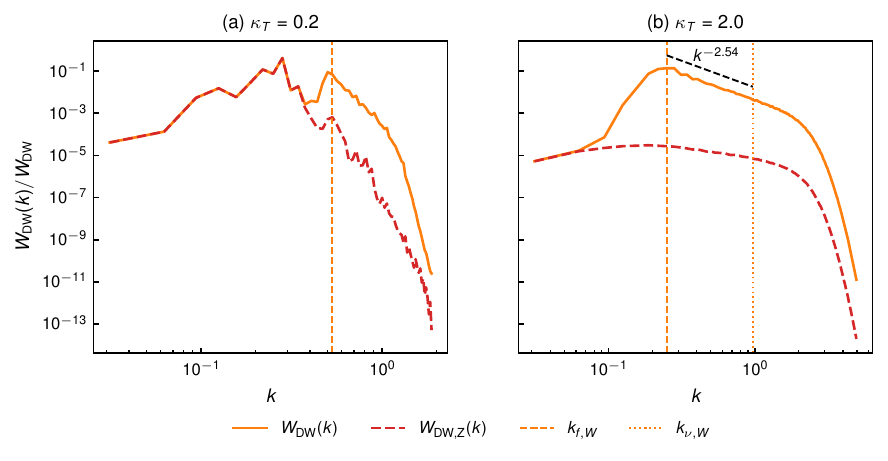}\caption{Normalized drift-wave potential enstrophy spectrum, $W_{\mathrm{DW}}(k)/W_{\mathrm{DW}}$,
vs. the perpendicular wavenumber, $k$, for PV-ITG. Dashed red curves
are the zonal component of the spectra. (a) For $\kappa_{T}=0.2$,
the non-zonal spectrum peaks after the zonal spectrum, just like the
total energy spectrum. (b) For $\kappa_{T}=2.0$, the spectrum scales
as $W_{\mathrm{DW}}\sim k^{-2.54}$ in the domain $(k_{f,W},k_{\nu,W})$
where $k_{f,W}$ and $k_{\nu,W}$ are the forcing and hyperviscous
scales of the potential enstrophy.\label{fig:WDW_comp_kapt_PV}}
\end{figure*}

Note that the drift-wave potential enstrophy is not a conserved quantity
for both the ST-ITG and PV-ITG systems due to the diamagnetic nonlinearity.
The spectra, time averaged over the second half of the simulation,
are shown in figure~\ref{fig:WDW_comp_kapt_PV}. For $\kappa_{T}=0.2$,
the non-zonal spectrum peaks after the zonal part, just as in the
total energy spectrum. For $\kappa_{T}=2.0$, the slope of the PV-ITG
spectrum in the domain $(k_{f,W},k_{\nu,W})$ is $-2.54$. 

\subsection{Spectrum of $W$ for PV-ITG}

The instantaneous spectrum of potential enstrophy is defined as
\begin{equation}
W(k,t)=\frac{1}{2\Delta k}\sum_{k-\Delta k/2<\left|\boldsymbol{q}\right|\le k+\Delta k/2}|-q^{2}\phi_{\boldsymbol{q}}-\tau\widetilde{\phi}_{\boldsymbol{q}}+P_{\boldsymbol{q}}/\Gamma|^{2}\,\text{.}
\end{equation}

\begin{figure*}
\centering{}\includegraphics{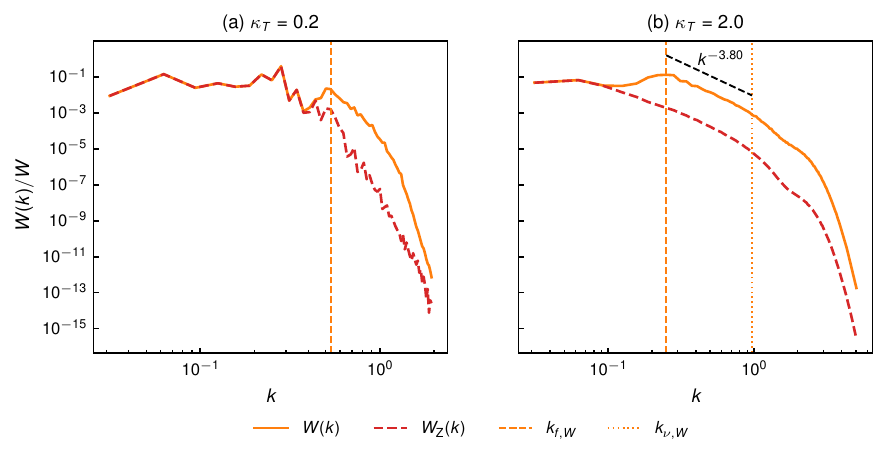}\caption{Normalized potential enstrophy spectrum, $W(k)/W$, vs. the perpendicular
wavenumber, $k$, for the PV-ITG model. (a) For $\kappa_{T}=0.2$,
the spectrum is like the drift-wave potential enstrophy spectrum and
has two peaks just like the total energy spectrum. (b) For $\kappa_{T}=2.0$,
the spectrum scales as $W(k)\sim k^{-3.80}$ in the domain $(k_{f,W},k_{\nu,W})$.\label{fig:Wk_comp_kapt}}
\end{figure*}

The spectra, time averaged over the second half of the simulation,
are shown in figure~\ref{fig:Wk_comp_kapt}. There is no scale separation
between $k_{f,W}$ and $k_{\nu,W}$ for $\kappa_{T}=0.2$. For $\kappa_{T}=0.2$,
the spectrum is quite similar to the drift-wave potential enstrophy
spectrum as shown in figure~\ref{fig:WDW_comp_kapt_PV}. However,
the $\kappa_{T}=2.0$ spectrum is steeper than that of the drift-wave
potential enstrophy spectrum {[}figure~\ref{fig:WDW_comp_kapt_PV}{]}
due to the contribution from the pressure and cross-term spectra resulting
in an exponent of $-3.80$ between $k_{f,W}$ and $k_{\nu,W}$. Furthermore,
the spectrum is flatter for $k<k_{f,W}$ compared to the drift-wave
potential enstrophy spectrum.

\section{Spectral flux of $E_{P}$, $G_{\mathrm{tot}}$ and $W$\label{appsec:spectral-fluxes}}

\subsection{Spectral flux of $E_{P}$ for ST-ITG}

The time evolution of $E_{P}(k,t)$ is
\begin{equation}
\partial_{t}E_{P}(k,t)+\partial_{k}\Pi_{P}(k,t)=f_{P}(k,t)-d^{(\nu)}_{P}(k,t)-d^{(H)}_{P}(k,t)\,\text{,}
\end{equation}

\noindent where the instantaneous spectral flux of $E_{P}$ through
wavenumber $k$ is
\begin{equation}
\Pi_{P}(k,t)=\Re\left(\sum_{\boldsymbol{q}\le k}P^{*}_{\boldsymbol{q}}\{\phi,P\}_{\boldsymbol{q}}\right)\,\text{.}
\end{equation}

Note that there is no FLR component in the $E_{P}(k,t)$ flux. The
instantaneous spectral production, spectral hyperviscous dissipation,
and spectral hypoviscous dissipation of $E_{P}$ are
\begin{align}
f_{P}(k,t) & =\frac{1}{\Delta k}\Re\left(\sum_{k<\left|\boldsymbol{q}\right|\le k+\Delta k}(\kappa_{n}+\kappa_{T})iq_{y}\phi^{*}_{\boldsymbol{q}}P_{\boldsymbol{q}}\right)\,\text{,}\\
d^{(\nu)}_{P}(k,t) & =\frac{1}{\Delta k}\sum_{k<|\boldsymbol{q}|\le k+\Delta k}\nu q^{6}\left|P_{\boldsymbol{q}}\right|^{2}\,\text{,}\\
d^{(H)}_{P}(k,t) & =\frac{1}{\Delta k}\sum_{k<\left|\boldsymbol{q}\right|\le k+\Delta k}Hq^{-4}\left|\widetilde{P}_{\boldsymbol{q}}\right|^{2}\,\text{.}
\end{align}

\begin{figure*}
\centering{}\includegraphics{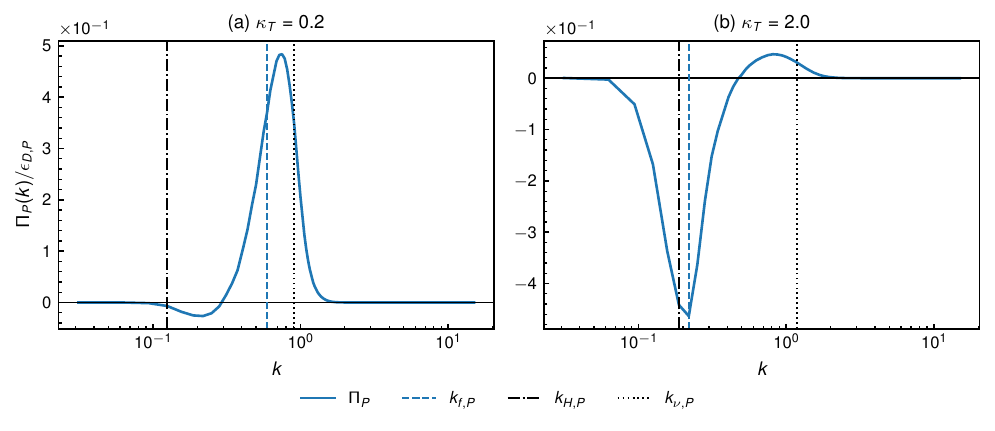}\caption{Spectral flux of $E_{P}$, $\Pi_{P}(k)$. (a) $\kappa_{T}=0.2$: Mostly
positive with a small domain where the flux is negative. (b) $\kappa_{T}=2.0$:
Mostly negative with a small domain where it is positive.\label{fig:EPk_flux_comp_kapt}}
\end{figure*}

The spectral flux of $E_{P}$, time averaged over the second half
of the simulation and normalized by the dissipation, $\epsilon_{D,P}$,
is shown for $\kappa_{T}=0.2$ and $\kappa_{T}=2.0$ in figure~\ref{fig:EPk_flux_comp_kapt}.
For $\kappa_{T}=0.2$, the transfer is mostly forward except for a
small domain, $(0.1,0.3)$, where the flux is significant and negative.
Whereas for $\kappa_{T}=2.0$ the transfer is predominantly negative
for medium wavenumbers except for a small domain near $k=1$ where
the flux reaches significant but small positive values. Furthermore,
the forcing scale, $k_{f,P}$, reduces with an increase in $\kappa_{T}$
just like the total energy.

\subsection{Spectral flux of $G_{\mathrm{tot}}$ for ST-ITG}

The time evolution of $G_{\mathrm{tot}}(k,t)$ is
\begin{equation}
\partial_{t}G_{\mathrm{tot}}(k,t)+\partial_{k}\Pi_{G}(k,t)=f_{G}(k,t)-d^{(\nu)}_{G}(k,t)-d^{(H)}_{G}(k,t)\,\text{,}
\end{equation}

\noindent where the spectral flux of $G_{\mathrm{tot}}$ through wavenumber
$k$ is
\begin{equation}
\begin{aligned}\Pi_{G}(k,t) & =\Re\left(\sum_{\boldsymbol{q}\le k}\phi^{*}_{\boldsymbol{q}}\{\phi,(\tau\widetilde{\phi}-\nabla^{2}\phi)\}_{\boldsymbol{q}}+P^{*}_{\boldsymbol{q}}\{\phi,(\tau\widetilde{\phi}-\nabla^{2}\phi)\}_{\boldsymbol{q}}\right)\\
 & +\Re\left(\sum_{\boldsymbol{q}\le k}(\mathds{1}_{q_{y}\neq0}+q^{2})\phi^{*}_{\boldsymbol{q}}\{\phi,P\}_{\boldsymbol{q}}+(\tau+q^{2})P^{*}_{\boldsymbol{q}}\{\phi,P\}_{\boldsymbol{q}}\right)\\
 & +\Re\left(\sum_{\boldsymbol{q}\le k}\phi^{*}_{\boldsymbol{q}}[\boldsymbol{\nabla}\cdot\{\boldsymbol{\nabla}\phi,P\}]_{\boldsymbol{q}}+P^{*}_{\boldsymbol{q}}[\boldsymbol{\nabla}\cdot\{\boldsymbol{\nabla}\phi,P\}]_{\boldsymbol{q}}\right)\\
 & =\Pi^{(\phi)}_{G}(k,t)+\Pi^{(d)}_{G}(k,t)\,\text{.}
\end{aligned}
\end{equation}

Here, $\Pi^{(\phi)}_{G}(k,t)$ is the instantaneous flux due to the
$\mathrm{E}\times\mathrm{B}$ nonlinearities (first four terms) and
$\Pi^{(d)}_{G}(k,t)$ is the instantaneous flux due to the diamagnetic
nonlinearity (last two terms). Note that the first term of $\Pi^{(\phi)}_{G}(k,t)$
and the first term of $\Pi^{(d)}_{G}(k,t)$ are simply $\Pi^{(\phi)}_{E}(k,t)$
and $\Pi^{(d)}_{E}(k,t)$ respectively {[}equation~(\ref{eq:spectral_flux_E}){]}.
The instantaneous spectral production, spectral hyperviscous dissipation,
and spectral hypoviscous dissipation of $G_{\mathrm{tot}}$ are
\begin{align}
f_{G}(k,t) & =\frac{1}{\Delta k}\Re\left(\sum_{k<\left|\boldsymbol{q}\right|\le k+\Delta k}[\kappa_{B}+(1+\tau)\kappa_{n}+\tau\kappa_{T}]iq_{y}\phi^{*}_{\boldsymbol{q}}P_{\boldsymbol{q}}\right)\,\text{,}\\
d^{(\nu)}_{G}(k,t) & =\frac{1}{\Delta k}\sum_{k<\left|\boldsymbol{q}\right|\le k+\Delta k}\nu q^{6}[\tau|\widetilde{\phi}_{\boldsymbol{q}}+P_{\boldsymbol{q}}|^{2}+q^{2}|\phi_{\boldsymbol{q}}+P_{\boldsymbol{q}}|^{2}]\,\text{,}\\
d^{(H)}_{G}(k,t) & =\frac{1}{\Delta k}\sum_{k<\left|\boldsymbol{q}\right|\le k+\Delta k}Hq^{-4}(\tau+q^{2})|\widetilde{\phi}_{\boldsymbol{q}}+\widetilde{P}_{\boldsymbol{q}}|^{2}\,\text{.}
\end{align}

\begin{figure*}
\centering{}\includegraphics{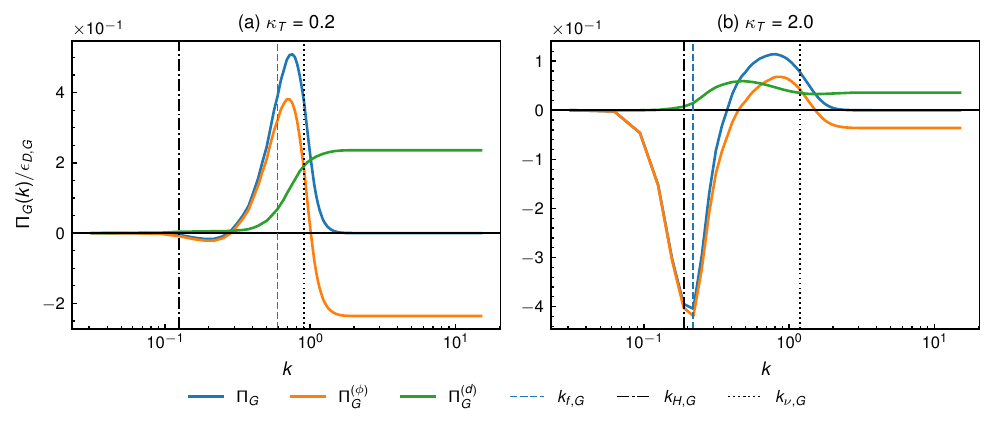}\caption{Spectral flux of $G_{\mathrm{tot}}$, $\Pi_{G}(k)$. (a) $\kappa_{T}=0.2$:
Predominantly positive. (b) $\kappa_{T}=2.0$: Predominantly negative
with a lower $k_{f,G}$. Furthermore, the components themselves do
not go to zero at large wavenumbers for both gradients.\label{fig:Gk_flux_comp_kapt}}
\end{figure*}

The spectral flux of $G_{\mathrm{tot}}$, time averaged over the second
half of the simulation and normalized by the dissipation, $\epsilon_{D,G}$,
is shown for $\kappa_{T}=0.2$ and $\kappa_{T}=2.0$ in figure~\ref{fig:Gk_flux_comp_kapt}.
The $\kappa_{T}=0.2$ flux is predominantly positive while the $\kappa_{T}=2.0$
flux is predominantly negative. Furthermore, the forcing scale, $k_{f,G}$,
reduces with an increase in $\kappa_{T}$ just like the other conserved
quantities. The $\mathrm{E}\times\mathrm{B}$ component and the diamagnetic
component do not go to zero at $k_{\mathrm{max}}$ on their own even
though their sum goes to zero. In addition, the diamagnetic component
is always positive and the shape of the flux for $k<k_{\nu,G}$ is
mostly determined by the $\mathrm{E}\times\mathrm{B}$ component for
both gradients. This reduced prominence of the diamagnetic component,
especially for $\kappa_{T}=2.0$, is different from its behavior in
the case of spectral flux of the total energy.

\subsection{Spectral flux of $W$ for PV-ITG}

The time evolution of $W(k,t)$ is
\begin{equation}
\partial_{t}W(k,t)+\partial_{k}\Pi_{W}(k,t)=f_{W}(k,t)-d^{(\nu)}_{W}(k,t)-d^{(H)}_{W}(k,t)\,\text{,}
\end{equation}

\noindent where the spectral flux of $W$ through wavenumber $k$
is
\begin{equation}
\Pi_{W}(k,t)=\Re\left(\sum_{\left|\boldsymbol{q}\right|\le k}(-q^{2}\phi^{*}_{\boldsymbol{q}}-\tau\widetilde{\phi}^{*}_{\boldsymbol{q}}+P^{*}_{\boldsymbol{q}}/\Gamma)\{\phi,\nabla^{2}\phi-\tau\widetilde{\phi}+P/\Gamma\}_{\boldsymbol{q}}\right)\,\text{.}
\end{equation}

Here, $\nabla^{2}\phi-\tau\widetilde{\phi}+P/\Gamma$ is the potential
vorticity. The spectral production, spectral hyperviscous dissipation,
and spectral hypoviscous dissipation of $W$ are
\begin{align}
f_{W}(k,t) & =\frac{1}{\Delta k}\Re\left(\sum_{k<\left|\boldsymbol{q}\right|\le k+\Delta k}[-\kappa_{n}+(\kappa_{n}+\kappa_{T})/\Gamma]/\Gamma iq_{y}\phi^{*}_{\boldsymbol{q}}P_{\boldsymbol{q}}\right)\,\text{,}\\
d^{(\nu)}_{W}(k,t) & =\frac{1}{\Delta k}\sum_{k<\left|\boldsymbol{q}\right|\le k+\Delta k}\nu q^{6}|-q^{2}\phi_{\boldsymbol{q}}-\tau\widetilde{\phi}_{\boldsymbol{q}}+P_{\boldsymbol{q}}/\Gamma|^{2}\,\text{,}\\
d^{(H)}_{W}(k,t) & =\frac{1}{\Delta k}\sum_{k<\left|\boldsymbol{q}\right|\le k+\Delta k}Hq^{-4}|-q^{2}\widetilde{\phi}_{\boldsymbol{q}}-\tau\widetilde{\phi}_{\boldsymbol{q}}+\widetilde{P}_{\boldsymbol{q}}/\Gamma|^{2}\,\text{.}
\end{align}

\begin{figure*}
\centering{}\includegraphics{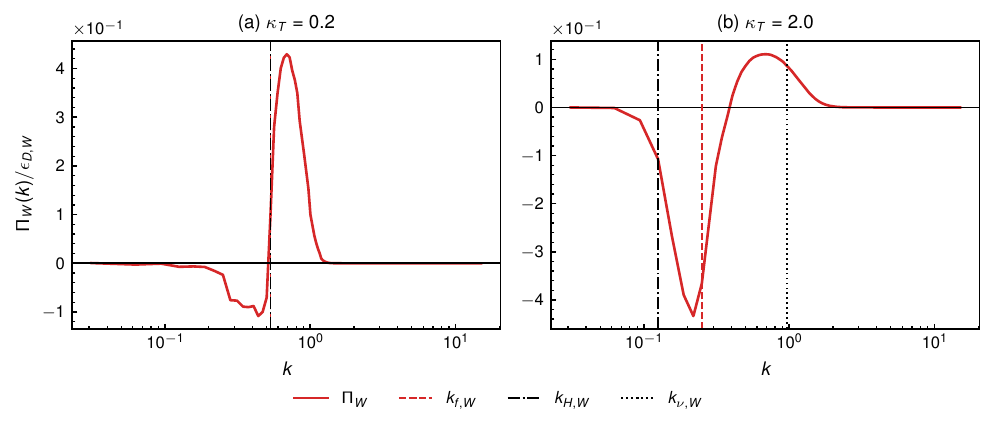}\caption{Spectral flux of $W$, $\Pi_{W}(k)$. (a) $\kappa_{T}=0.2$: Minimal
scale separation between $k_{H,W}$, $k_{f,W}$ and $k_{\nu,W}$.
Mostly forward transfer for $k>k_{f,W}$. (b) $\kappa_{T}=2.0$: Prominent
inverse transfer of potential enstrophy for intermediate scales and
higher scale separation between $k_{H,W}$, $k_{f,W}$ and $k_{\nu,W}$.\label{fig:Wk_flux_comp_kapt}}
\end{figure*}

The spectral flux, averaged over the second half of the simulation,
$\Pi_{W}(k)$, and the scales corresponding to forcing and dissipation
are shown in figure~\ref{fig:Wk_flux_comp_kapt}. For $\kappa_{T}=0.2$,
there is limited scale separation and the flux is negative below $k_{f,W}$
and positive above $k_{f,W}$. The positive maximum is larger in magnitude
than the negative minimum. The $\kappa_{T}=2.0$ case displays higher
scale separation between the hypoviscous and hyperviscous scales and
a dominant inverse transfer of potential enstrophy at intermediate
scales. Note that the transfer of total energy in a limited domain
around $k_{f,W}$ is towards higher wavenumbers resulting in the total
energy and potential enstrophy cascading in opposite directions. However
for $k>0.4$, both the total energy and potential enstrophy are nonlinearly
transferred to higher wavenumbers. The inverse transfer of potential
enstrophy observed for $\kappa_{T}=2.0$, unlike what is seen in 2D
Navier-Stokes, is facilitated by the presence of pressure terms and
cross-terms in the conservation budget of potential enstrophy.

\section{Grid convergence\label{appsec:spectra-1}}

We perform grid-convergence tests for both ST-ITG and PV-ITG at $\kappa_{T}=2.0$,
using padded spectral resolutions $N^{2}_{p}=512^{2},768^{2},$ and
$1024^{2}$, while keeping the box size and all model parameters fixed.
We choose $\kappa_{T}=2.0$ because it exhibits strong turbulence
in contrast to the zonal-flow-dominated regime at $\kappa_{T}=0.2$,
and is therefore expected to provide a stringent test of the numerical
resolution. Simulations at resolutions beyond $N^{2}_{p}=1024^{2}$
were not performed due to their computational cost. The grid-convergence
results are summarized in table~\ref{tab:grid-convergence}, where
we compare the time-averaged zonal kinetic energy fraction ($\langle\Xi\rangle_{\Delta t}$)
and box-averaged heat flux ($\langle Q_{\mathrm{box}}\rangle_{\Delta t}$)
for both models. The time averages are computed over the interval
$\gamma t\in[200,400]$. 

The relative changes in $\langle\Xi\rangle_{\Delta t}$ between successive
resolutions remain below $10\%$ and $5\%$ for ST-ITG and PV-ITG
respectively. For ST-ITG, the time-averaged zonal kinetic energy fraction
is highest at $N^{2}_{p}=768^{2}$, whereas it increases monotonically
with resolution for PV-ITG. Similarly, the relative changes in $\langle Q_{\mathrm{box}}\rangle_{\Delta t}$
remain below $15\%$ and $20\%$ for ST-ITG and PV-ITG respectively.
The relative changes in $\langle Q_{\mathrm{box}}\rangle_{\Delta t}$
exhibit a somewhat greater sensitivity to resolution than $\langle\Xi\rangle_{\Delta t}$.
However, note that $\langle Q_{\mathrm{box}}\rangle_{\Delta t}$ is
the lowest at $N^{2}_{p}=768^{2}$ for both models.

\begin{table}
\begin{centering}
\begin{tabular}{|c|c|c|c|c|c|}
\hline 
Model & $N^{2}_{p}$ & $\langle\Xi\rangle_{\Delta t}$ & $\langle Q_{\mathrm{box}}\rangle_{\Delta t}$ & $\Delta\langle\Xi\rangle_{\Delta t}(\%)$ & $\Delta\langle Q_{\mathrm{box}}\rangle_{\Delta t}(\%)$\tabularnewline
\hline 
\hline 
\multirow{3}{*}{ST-ITG} & $512^{2}$ & $0.080$ & $129.207$ & N/A & N/A\tabularnewline
\cline{2-6}
 & $768^{2}$ & $0.088$ & $125.545$ & $+9.11\%$ & $-2.83\%$\tabularnewline
\cline{2-6}
 & $1024^{2}$ & $0.087$ & $142.865$ & $-0.57\%$ & $+13.80\%$\tabularnewline
\hline 
\multirow{3}{*}{PV-ITG} & $512^{2}$ & $0.145$ & $47.895$ & N/A & N/A\tabularnewline
\cline{2-6}
 & $768^{2}$ & $0.146$ & $42.503$ & $+0.67\%$ & $-11.26\%$\tabularnewline
\cline{2-6}
 & $1024^{2}$ & $0.153$ & $50.161$ & $+4.86\%$ & $+18.02\%$\tabularnewline
\hline 
\end{tabular}
\par\end{centering}
\caption{Grid-convergence tests for ST-ITG and PV-ITG at $\kappa_{T}=2.0$.
The time-averaged zonal kinetic energy fraction ($\langle\Xi\rangle_{\Delta t}$)
and box-averaged heat flux ($\langle Q_{\mathrm{box}}\rangle_{\Delta t}$)
are reported for three padded resolutions. The relative changes $\Delta\langle\Xi\rangle_{\Delta t}$
and $\Delta\langle Q_{\mathrm{box}}\rangle_{\Delta t}$ are calculated
between successive resolutions.\label{tab:grid-convergence}}
\end{table}

\begin{figure*}
\centering{}\includegraphics{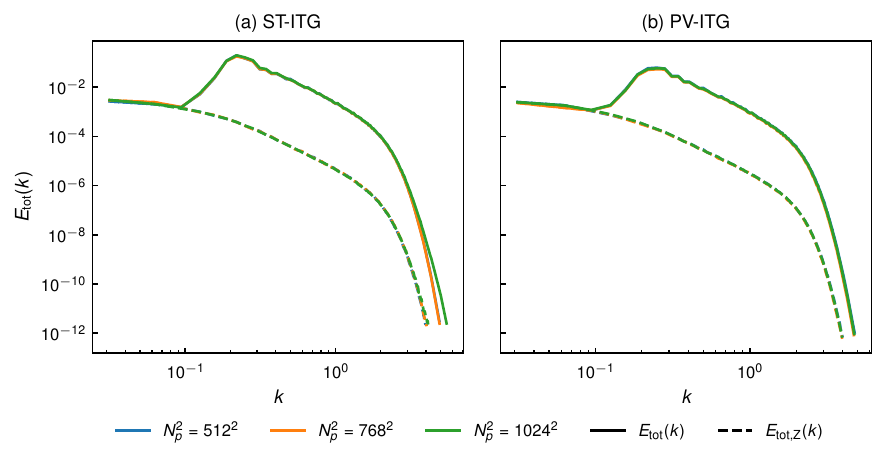}\caption{Grid-convergence test of the total energy spectrum, $E_{\mathrm{tot}}(k)$,
as a function of the perpendicular wavenumber, $k$, at $\kappa_{T}=2.0$.
The spectra are compared for padded spectral resolutions $N^{2}_{p}=512^{2}$,
$768^{2}$, and $1024^{2}$. The dashed curves correspond to the zonal
component of the spectra. (a) ST-ITG: the spectra show good agreement,
except at high wavenumbers, where the $512^{2}$ and $768^{2}$ spectra
fall off faster than the $1024^{2}$ spectrum. (b) PV-ITG: the spectra
show no visible difference across the three resolutions.\label{fig:Ek_comp_convergence}}
\end{figure*}

We further examine spectral convergence by comparing the total energy
spectra, $E_{\mathrm{tot}}(k)$, at the three resolutions, as shown
in figure~\ref{fig:Ek_comp_convergence}. The spectra are obtained
by time-averaging the instantaneous total energy spectra over the
interval $\gamma t\in[200,400]$. For both models, the spectra and
their zonal components show good agreement across all three resolutions.
The only visible difference is that for ST-ITG the spectra fall off
faster for $512^{2}$ and $768^{2}$ than for $1024^{2}$. This indicates
that the spectral properties are relatively insensitive to numerical
resolution over the range considered. The strong hyperviscous damping
of small-scale modes likely contributes to this weak resolution dependence.

\end{document}